\documentclass[%
 reprint,
superscriptaddress,
showkeys,
 amsmath,amssymb,
 aps,
prb,
floatfix
]{revtex4-2}

\usepackage{hyperref,graphicx}% Include figure files
\usepackage{dcolumn}% Align table columns on decimal point
\usepackage{bm}% bold math
\usepackage{textgreek}
\usepackage[utf8]{inputenc}

\begin{document}
%%%%%%%%%%%%%%%%%%%%%%%%%%%%%%%%%%%%%%%%%%%%%%%%%%%%%%%%%%%%%%%%%%%%%%%%%%%%%%%%%%%%%%%%%%%%%%%%%%%%%%%%%%%%
%Authors
%%%%%%%%%%%%%%%%%%%%%%%%%%%%%%%%%%%%%%%%%%%%%%%%%%%%%%%%%%%%%%%%%%%%%%%%%%%%%%%%%%%%%%%%%%%%%%%%%%%%%%%%%%%%
\title{Interplay between local moment and itinerant magnetism in the layered metallic antiferromagnet TaFe\textsubscript{1.14}Te\textsubscript{3}}

\author{Sae Young Han}
\affiliation{Department of Chemistry, Columbia University, New York, NY, USA}

\author{Evan J. Telford}%
\email{ejt2133@columbia.edu}
\affiliation{Department of Chemistry, Columbia University, New York, NY, USA}
\affiliation{Department of Physics, Columbia University, New York, NY, USA}

\author{Asish K. Kundu}
\affiliation{Condensed Matter Physics and Materials Science Department, Brookhaven National Laboratory, Upton, NY, USA}

\author{Sylvia J. Bintrim}
\affiliation{Department of Chemistry, Columbia University, New York, NY, USA}

\author{Simon Turkel}
\affiliation{Department of Physics, Columbia University, New York, NY, USA}

\author{Ren A. Wiscons}
\affiliation{Department of Chemistry, Columbia University, New York, NY, USA}
\affiliation{Department of Chemistry, Amherst College, Amherst, MA, USA}

\author{Amirali Zangiabadi}
\affiliation{Department of Applied Physics and Applied Mathematics, Columbia University, New York, NY, USA}

\author{Eun-Sang Choi}
\affiliation{National High Magnetic Field Laboratory, Tallahassee, FL, USA}

\author{Tai-De Li}
\affiliation{Nanoscience Initiative at Advanced Science Research Center, Graduate Center of the City University of New York, New York, NY, USA}
\affiliation{Department of Physics, City College of New York, City University of New York, New York, NY, USA}

\author{Michael L. Steigerwald}
\affiliation{Department of Chemistry, Columbia University, New York, NY, USA}

\author{Timothy C. Berkelbach}
\affiliation{Department of Chemistry, Columbia University, New York, NY, USA}

\author{Abhay N. Pasupathy}
\affiliation{Condensed Matter Physics and Materials Science Department, Brookhaven National Laboratory, Upton, NY, USA}

\author{Cory R. Dean}
\affiliation{Department of Physics, Columbia University, New York, NY, USA}

\author{Colin Nuckolls}
\affiliation{Department of Chemistry, Columbia University, New York, NY, USA}

\author{Xavier Roy}
\email{xr2114@columbia.edu}
\affiliation{Department of Chemistry, Columbia University, New York, NY, USA}
\date{\today}% It is always \today, today,
%Abstract
%%%%%%%%%%%%%%%%%%%%%%%%%%%%%%%%%%%%%%%%%%%%%%%%%%%%%%%%%%%%%%%%%%%%%%%%%%%%%%%%%%%%%%%%%%%%%%%%%%%%%%%%%%%%
\begin{abstract}
Two-dimensional (2D) antiferromagnets have garnered considerable interest for the next generation of functional spintronics. However, many available bulk materials from which 2D antiferromagnets are isolated are limited by their sensitivity to air, low ordering temperatures, and insulating transport properties. TaFe\textsubscript{1+\textit{y}}Te\textsubscript{3} offers unique opportunities to address these challenges with increased air stability, metallic transport properties, and robust antiferromagnetic order. Here we synthesize TaFe\textsubscript{1+\textit{y}}Te\textsubscript{3} (\textit{y} = 0.14), identify its structural, magnetic, and electronic properties, and elucidate the relationships between them. Axial-dependent high-field magnetization measurements on TaFe\textsubscript{1.14}Te\textsubscript{3} reveal saturation magnetic fields ranging between 27-30 T with a saturation magnetic moment between 2.05-2.12 \textmu\textsubscript{B}. Magnetotransport measurements confirm TaFe\textsubscript{1.14}Te\textsubscript{3} is metallic with strong coupling between magnetic order and electronic transport. Angle-resolved photoemission spectroscopy measurements across the magnetic transition uncover a complex interplay between itinerant electrons and local magnetic moments that drives the magnetic transition. We further demonstrate the ability to isolate few-layer sheets of TaFe\textsubscript{1.14}Te\textsubscript{3} through mechanical exfoliation, establishing TaFe\textsubscript{1.14}Te\textsubscript{3} as a potential platform for two-dimensional spintronics based on metallic layered antiferromagnets.
\end{abstract}
%%%%%%%%%%%%%%%%%%%%%%%%%%%%%%%%%%%%%%%%%%%%%%%%%%%%%%%%%%%%%%%%%%%%%%%%%%%%%%%%%%%%%%%%%%%%%%%%%%%%%%%%%%%%
%Keywords
%%%%%%%%%%%%%%%%%%%%%%%%%%%%%%%%%%%%%%%%%%%%%%%%%%%%%%%%%%%%%%%%%%%%%%%%%%%%%%%%%%%%%%%%%%%%%%%%%%%%%%%%%%%%
%%%%%%%%%%%%%%%%%%%%%%%%%%%%%%%%%%%%%%%%%%%%%%%%%%%%%%%%%%%%%%%%%%%%%%%%%%%%%%%%%%%%%%%%%%%%%%%%%%%%%%%%%%%%
%Make the title
%%%%%%%%%%%%%%%%%%%%%%%%%%%%%%%%%%%%%%%%%%%%%%%%%%%%%%%%%%%%%%%%%%%%%%%%%%%%%%%%%%%%%%%%%%%%%%%%%%%%%%%%%%%%
\maketitle
%\tableofcontents
%%%%%%%%%%%%%%%%%%%%%%%%%%%%%%%%%%%%%%%%%%%%%%%%%%%%%%%%%%%%%%%%%%%%%%%%%%%%%%%%%%%%%%%%%%%%%%%%%%%%%%%%%%%%
%Introduction
%%%%%%%%%%%%%%%%%%%%%%%%%%%%%%%%%%%%%%%%%%%%%%%%%%%%%%%%%%%%%%%%%%%%%%%%%%%%%%%%%%%%%%%%%%%%%%%%%%%%%%%%%%%%
\section*{\label{sec:intro} Introduction}
Identifying candidate low-dimensional materials combining metallic transport behavior with robust antiferromagnetic order is crucial to advancing spin-based nanotechnologies. Van der Waals (vdW) magnets, in particular, provide an ideal platform for fabricating atomically-thin spintronics and magneto-optic devices\cite{RN1,RN2,RN3,RN4,RN5}, as they possess diverse magnetic phases and electronic properties that persist down to the two-dimensional (2D) limit. Within this family of materials, those exhibiting metallic electronic transport have received intense attention due to their potential functionality as spin injectors\cite{RN6,RN7}, magnetic tunnel junction electrodes\cite{RN8}, or spin-orbit torque devices\cite{RN9,RN10}. Among the available 2D magnets, layered antiferromagnets are exceptionally attractive due to the increased resilience to external magnetic fields\cite{RN11,RN12}, sensitivity of electrical resistance to field-induced antiferromagnetic-to-ferromagnetic transitions\cite{RN13,RN14}, faster spin dynamics\cite{RN15}, utility in high-frequency spin-orbit torque experiments\cite{RN16}, and capability for higher packing density than ferromagnets in memory storage devices due to the absence of stray fields\cite{RN17}. Despite an obvious need, vdW materials integrating metallic behavior with layered antiferromagnetic order are notably scarce\cite{RN18}. TaFe\textsubscript{1+\textit{y}}Te\textsubscript{3}\cite{RN19,RN20,RN21,RN22} is one candidate to overcome many existing limitations of vdW magnets, boasting a high Néel temperature ranging between 160 and 200 K\cite{RN19,RN21,RN23,RN24,RN25}, metallic charge transport properties\cite{RN19,RN23,RN24,RN26,RN27}, and stability under ambient conditions, showcasing its promise for metallic antiferromagnet-based spintronics. However, a comprehensive understanding of the magnetic and electronic properties of TaFe\textsubscript{1+\textit{y}}Te\textsubscript{3} has yet to be achieved.

%%%%%%%%%%%%%%%%%%%%%%%%%%%%%%%%%%%%%%%%%%%%%%%%%%%%%%%%%%%%%%%%%%%%%%%%%%%%%%%%%%%%%%%%%%%%%%%%%%%%%%%%%%%%
%Figure 1
%%%%%%%%%%%%%%%%%%%%%%%%%%%%%%%%%%%%%%%%%%%%%%%%%%%%%%%%%%%%%%%%%%%%%%%%%%%%%%%%%%%%%%%%%%%%%%%%%%%%%%%%%%%%
\begin{figure*}[t]
\includegraphics[width=6.9in]{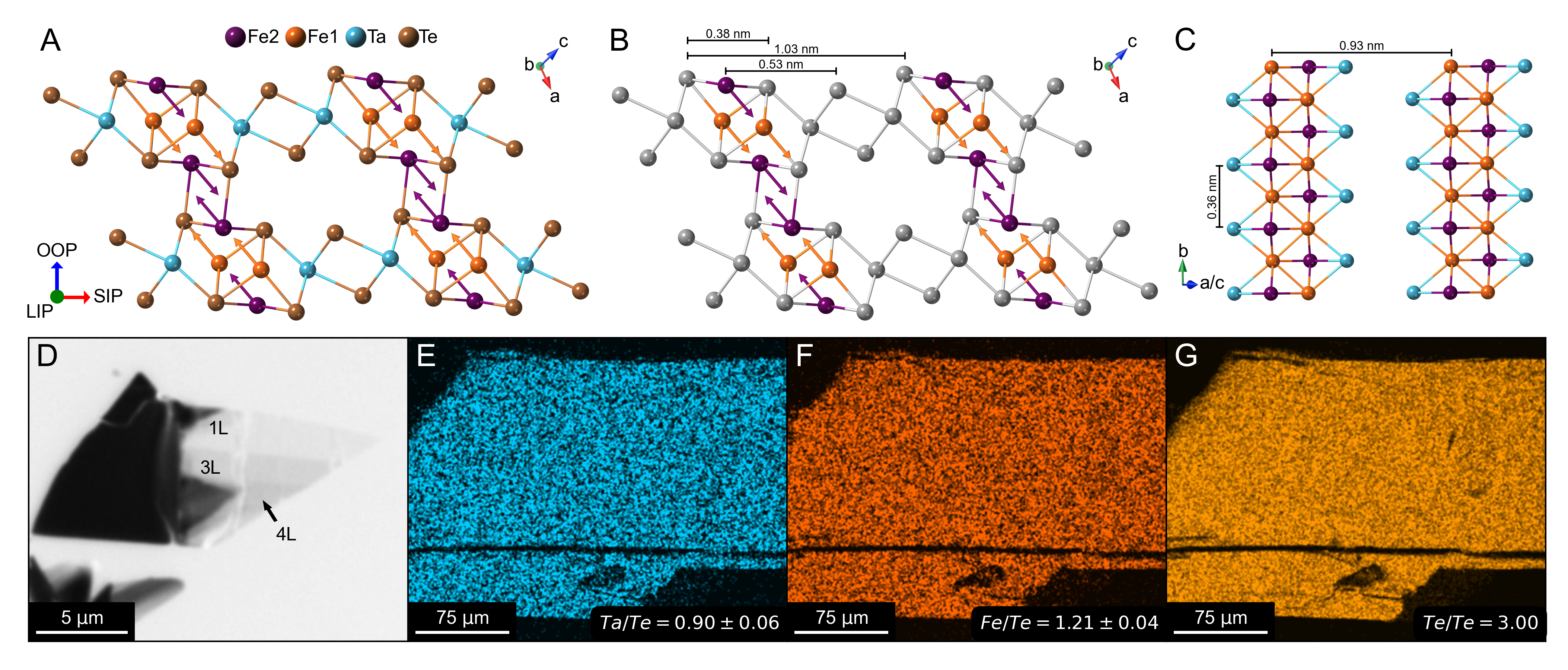}% Here is how to import EPS art
\centering
\caption[Structure and composition of TaFe\textsubscript{1.14}Te\textsubscript{3}]{\label{fig:figure1} Structure and composition of TaFe\textsubscript{1.14}Te\textsubscript{3}. A) Crystal structure of TaFe\textsubscript{1.14}Te\textsubscript{3} viewed along the crystallographic \textit{b}-axis. Morphological long in-plane (LIP) (green), short in-plane (SIP) (red), and out-of-plane (OOP) (blue) directions are shown relative to the crystallographic \textit{a}, \textit{b}, and \textit{c}-axes. B) Crystal structure of TaFe\textsubscript{1.14}Te\textsubscript{3} viewed along the crystallographic \textit{b}-axis with the orientation of the Fe1 (orange arrows) and Fe2 (purple arrows) magnetic moments in the antiferromagnetic state overlayed. C) Crystal structure of TaFe\textsubscript{1.14}Te\textsubscript{3} viewed along a direction that clearly displays the Fe1/Fe2 chain structure. The Te atoms are removed for clarity. D) False-colored optical microscope image of a TaFe\textsubscript{1.14}Te\textsubscript{3} flake exfoliated onto polydimethylsiloxane (PDMS). The corresponding layer numbers for different regions were determined by atomic force microscopy and optical contrast (Fig. \ref{fig:figures1}). E-G) Ta (E), Fe (F), and Te (G) elemental maps of a single crystal of TaFe\textsubscript{1.14}Te\textsubscript{3} as determined by SEM/EDX. The lower right inset of each elemental map gives the corresponding elemental composition normalized to Te. In (E) and (F), the error bars represent the standard deviation of multiple measurements between various crystals and growth batches.}
\end{figure*}
%%%%%%%%%%%%%%%%%%%%%%%%%%%%%%%%%%%%%%%%%%%%%%%%%%%%%%%%%%%%%%%%%%%%%%%%%%%%%%%%%%%%%%%%%%%%%%%%%%%%%%%%%%%%
%%%%%%%%%%%%%%%%%%%%%%%%%%%%%%%%%%%%%%%%%%%%%%%%%%%%%%%%%%%%%%%%%%%%%%%%%%%%%%%%%%%%%%%%%%%%%%%%%%%%%%%%%%%%

In this work, we synthesize TaFe\textsubscript{1+\textit{y}}Te\textsubscript{3} (\textit{y} = 0.14) and explore its electronic and magnetic properties, uncovering its potential for atomic-scale spintronic devices based on vdW materials. Temperature- and magnetic-field-dependent magnetization measurements reveal a paramagnetic (paramagnetic)-to-antiferromagnetic transition at 164 K followed by a transition into a spin-glassy magnetic phase below $\approx$100 K. At liquid helium temperature, full magnetization saturation occurs only at 27-30 T for all morphological directions with a saturation magnetization moment of 2.05-2.12 \textmu\textsubscript{B}. Through density functional theory (DFT) calculations, we confirm the stability of the metallic antiferromagnetic phase and characterize its electronic structure. Longitudinal magnetotransport and Hall measurements verify the metallic nature and demonstrate a strong coupling between magnetic order and electronic transport. Angle-resolved photoemission spectroscopy (ARPES) measurements further support the metallic nature of TaFe\textsubscript{1.14}Te\textsubscript{3} and reveal a complex interplay between local moment and itinerant magnetism. We also demonstrate the ability to isolate few-layer flakes of TaFe\textsubscript{1.14}Te\textsubscript{3} through mechanical exfoliation, solidifying its practicality for spintronic applications.
%%%%%%%%%%%%%%%%%%%%%%%%%%%%%%%%%%%%%%%%%%%%%%%%%%%%%%%%%%%%%%%%%%%%%%%%%%%%%%%%%%%%%%%%%%%%%%%%%%%%%%%%%%%%
%Results and Discussion
%%%%%%%%%%%%%%%%%%%%%%%%%%%%%%%%%%%%%%%%%%%%%%%%%%%%%%%%%%%%%%%%%%%%%%%%%%%%%%%%%%%%%%%%%%%%%%%%%%%%%%%%%%%%
\section*{\label{sec:results} Results and Discussion}
Single crystals of TaFe\textsubscript{1.14}Te\textsubscript{3} are grown using the chemical vapor transport method (see methods for full details). Tantalum, iron, tellurium, and TeCl\textsubscript{4} powders are pressed into a pellet, and sealed in a fused silica tube under vacuum. After seven days in a temperature gradient of 700°C to 600°C, needle-like crystals are obtained in the sink side of the tube. We performed single crystal X-ray diffraction (SCXRD) to determine the structure (Fig. \ref{fig:figure1}A-C). The low-dimensional layered telluride TaFe\textsubscript{1.14}Te\textsubscript{3} crystallizes in the \textit{P}2\textsubscript{1}/\textit{m} space group. The structure features Ta-Fe1-Fe1-Ta ribbons aligned along the \textit{b}-axis while sandwiched between Te layers. Along the ribbon, the Ta atoms are octahedrally coordinated by six Te atoms, and the Fe1 atoms, which make a one-dimensional zig-zag chain, are tetrahedrally coordinated by four Te atoms. Unlike the Fe1 site, the Fe2 site is partially occupied, and the Fe2 atom nestles in the center of the square pyramid formed by Te atoms (Fig. \ref{fig:figure1}A). Single crystals are stable under ambient conditions and the layered nature of the structure and corresponding interlayer vdW bonds allows for the isolation of air-stable few-layer flakes of TaFe\textsubscript{1.14}Te\textsubscript{3} through mechanical exfoliation down to the monolayer limit (Fig. \ref{fig:figure1}D). Corresponding tapping-mode atomic force microscopy images of an exfoliated sample can be found in Fig. \ref{fig:figures1}. Previously reported neutron diffraction measurements on TaFe\textsubscript{1+\textit{y}}Te\textsubscript{3} determined that the magnetic moments are localized on the Fe atoms and that the magnetic structure is A-type antiferromagnetic with Fe1 spins within each layer coupling ferromagnetically while Fe1 spins between the layers couple antiferromagnetically\cite{RN23}. The Fe2 spins within each layer align parallel with Fe1 spins (Fig. \ref{fig:figure1}B). The atomic ratio of different constituent elements was determined through SCXRD occupancy refinement (Tab. \ref{tab:tables1}) and confirmed by energy dispersive X-ray spectroscopy (EDX) on freshly cleaved samples (Tab. \ref{tab:tables2}). Corresponding scanning electron microscopy (SEM)/EDX elemental maps of a TaFe\textsubscript{1.14}Te\textsubscript{3} single crystal for Ta, Fe, and Te with each collection normalized to Te demonstrate that there is no evidence of Fe clustering down to the micron scale (Fig. \ref{fig:figure1}E-G). 

\begin{figure}[t]
\includegraphics[width=3.4in]{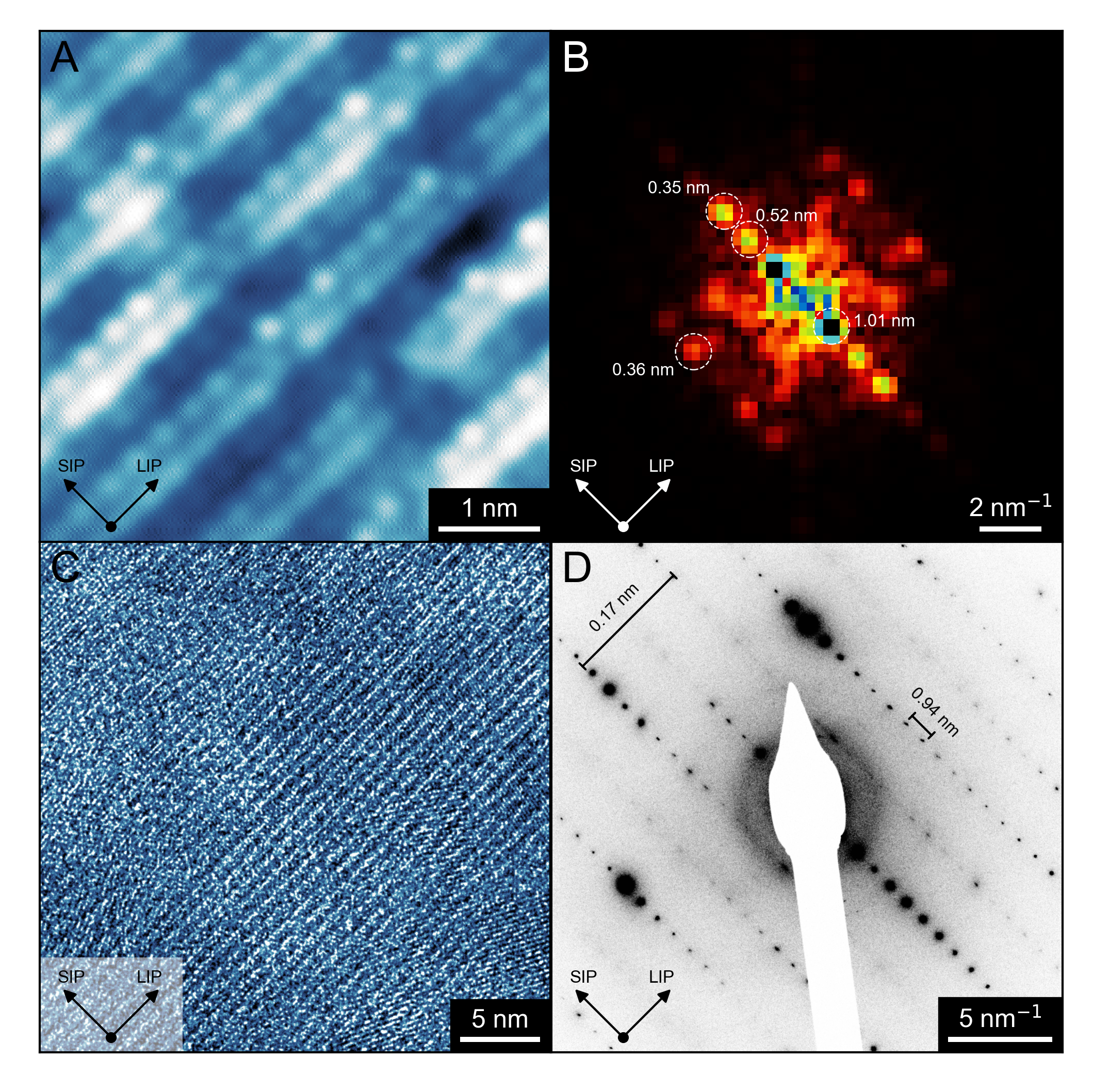}
\centering
\caption[Atomic-scale characterization of TaFe\textsubscript{1.14}Te\textsubscript{3}]{\label{fig:figure2} Atomic-scale characterization of TaFe\textsubscript{1.14}Te\textsubscript{3}. A) STM image of TaFe\textsubscript{1.14}Te\textsubscript{3} cleaved along the vdW gap. The image was obtained at \textit{T} = 7.5 K in constant current mode (\textit{V}\textsubscript{b} = 400 mV, \textit{I} = 50 pA). B) Corresponding FFT of the STM image. Select periodicities are denoted by dashed white circles. C) TEM micrograph of an exfoliated TaFe\textsubscript{1.14}Te\textsubscript{3} flake with a thickness $<$ 50 nm viewed along the OOP direction. D) Corresponding electron diffraction image of the same flake in (C). Select \textit{d}-spacings are denoted by solid black bars. In all panels, the orientation of the images with respect to the morphological SIP and LIP directions is given in the lower left inset.}
\end{figure}

Since the Fe2 site is only partially occupied, we first characterize the uniformity of the Fe2 distribution and the impact of Fe2 on the local electronic properties. We performed scanning tunneling microscopy (STM) and spectroscopy (STS) on cleaved bulk TaFe\textsubscript{1.14}Te\textsubscript{3}. In Fig. \ref{fig:figure2}A, we present an atomically-resolved STM image of the vdW plane at 7.5 K. The selected distances determined by a fast Fourier transform (FFT) of the real-space STM topography are Te $\cdots$ Te (nearest neighbor) = 0.35, 0.36 nm, Fe2 $\cdots$ Te = 0.52 nm, and Te $\cdots$ Te (next nearest neighbor) = 1.01 nm (Fig. \ref{fig:figure2}B), in good agreement with those obtained from SCXRD (Fig. \ref{fig:figure1}A). With its atomic precision, STM allows us to view the topographical corrugation of the Fe chains in detail. We observe an inhomogeneity of the intensity in the image, which may indicate random nanometer-scale clustering of Fe2 in the crystal, as previously hypothesized in a Mössbauer spectroscopy study of TaFe\textsubscript{1+\textit{y}}Te\textsubscript{3}\cite{RN20}. Corresponding STS measurements show a metallic-like density of states (DOS) at all points across multiple Fe chains, revealing that the occupancy of the Fe2 site does not significantly impact the local electronic structure (Fig. \ref{fig:figures2}).

%%%%%%%%%%%%%%%%%%%%%%%%%%%%%%%%%%%%%%%%%%%%%%%%%%%%%%%%%%%%%%%%%%%%%%%%%%%%%%%%%%%%%%%%%%%%%%%%%%%%%%%%%%%%
%Figure 2
%%%%%%%%%%%%%%%%%%%%%%%%%%%%%%%%%%%%%%%%%%%%%%%%%%%%%%%%%%%%%%%%%%%%%%%%%%%%%%%%%%%%%%%%%%%%%%%%%%%%%%%%%%%%
%%%%%%%%%%%%%%%%%%%%%%%%%%%%%%%%%%%%%%%%%%%%%%%%%%%%%%%%%%%%%%%%%%%%%%%%%%%%%%%%%%%%%%%%%%%%%%%%%%%%%%%%%%%%
%%%%%%%%%%%%%%%%%%%%%%%%%%%%%%%%%%%%%%%%%%%%%%%%%%%%%%%%%%%%%%%%%%%%%%%%%%%%%%%%%%%%%%%%%%%%%%%%%%%%%%%%%%%%

To confirm that the quasi-one-dimensional nature of the structure persists in exfoliated flakes, we performed transmission electron microscopy (TEM) and corresponding selected-area electron diffraction (SAED) measurements on an exfoliated flake of TaFe\textsubscript{1.14}Te\textsubscript{3} with a thickness $<$ 50 nm (Fig. \ref{fig:figure2}C, D; see methods for details). The diffraction pattern shows that the crystallinity is intact in the exfoliated flake of TaFe\textsubscript{1.14}Te\textsubscript{3}. This is also seen in the real-space image, in which we observe distinct lines of atoms along the direction of the chains as depicted in Fig. \ref{fig:figure1}C. Select atomic distances from the SAED image are 0.17 nm for the close-range Fe atoms and 0.94 nm for the interchain Fe atoms (Fig. \ref{fig:figure2}D), in good agreement with those obtained from SCXRD (Tab. \ref{tab:tables1} and Fig. \ref{fig:figure1}A, B) and STM (Fig. \ref{fig:figure2}A, B). 

%%%%%%%%%%%%%%%%%%%%%%%%%%%%%%%%%%%%%%%%%%%%%%%%%%%%%%%%%%%%%%%%%%%%%%%%%%%%%%%%%%%%%%%%%%%%%%%%%%%%%%%%%%%%
%Figure 3
%%%%%%%%%%%%%%%%%%%%%%%%%%%%%%%%%%%%%%%%%%%%%%%%%%%%%%%%%%%%%%%%%%%%%%%%%%%%%%%%%%%%%%%%%%%%%%%%%%%%%%%%%%%%
\begin{figure*}[t]
\includegraphics[width=6.9in]{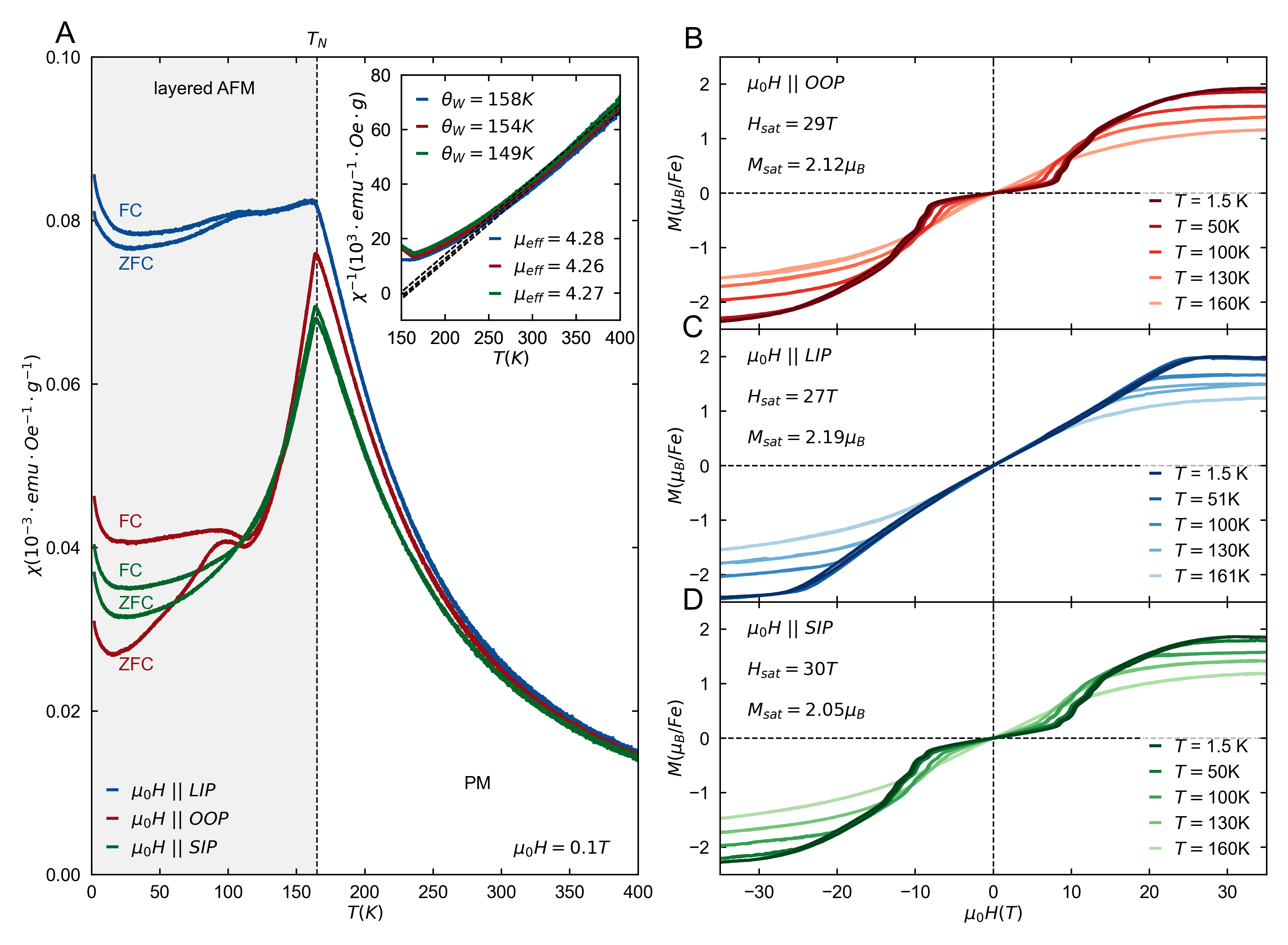}% Here is how to import EPS art
\centering
\caption[High-field magnetic properties of TaFe\textsubscript{1.14}Te\textsubscript{3}]{\label{fig:figure3} High-field magnetic properties of TaFe\textsubscript{1.14}Te\textsubscript{3}. A) \textchi{ }versus \textit{T} for \textit{H} oriented along the SIP (solid green lines), LIP (solid blue lines), and OOP (solid red lines) directions. Both zero-field-cooled (ZFC) and field-cooled (FC) traces are shown. All traces were collected with a field of \textit{H} = 0.1 T. The extracted Néel temperature is shown by a black dashed line. Layered antiferromagnetic and paramagnetic phases are denoted by grey and white regions, respectively. The inset shows inverse \textchi{ }versus \textit{T} for \textit{H} oriented along the SIP (solid green line), LIP (solid blue line), and OOP (solid red line) directions. Dashed black lines are linear fits to each trace. The extracted \texttheta\textsubscript{w} and \textmu\textsubscript{eff} are given in the upper left and lower right insets, respectively. B-D) \textit{M} versus \textit{H} at different \textit{T}s for \textit{H} oriented along the OOP (B), LIP (C), and SIP (D) directions. In each plot, the extracted saturation magnetization (\textit{M}\textsubscript{sat}) and \textit{H}\textsubscript{sat} are given in the inset. Note that because both the OOP and SIP directions contain a component parallel to the Fe1/Fe2 spin directions, they both exhibit metamagnetic transitions. The LIP direction, however, is completely perpendicular to the Fe1/Fe2 spin directions and only displays a gradual spin-canting process.}
\end{figure*}
%%%%%%%%%%%%%%%%%%%%%%%%%%%%%%%%%%%%%%%%%%%%%%%%%%%%%%%%%%%%%%%%%%%%%%%%%%%%%%%%%%%%%%%%%%%%%%%%%%%%%%%%%%%%
%%%%%%%%%%%%%%%%%%%%%%%%%%%%%%%%%%%%%%%%%%%%%%%%%%%%%%%%%%%%%%%%%%%%%%%%%%%%%%%%%%%%%%%%%%%%%%%%%%%%%%%%%%%%

We measured the magnetic susceptibility (\textchi) of TaFe\textsubscript{1.14}Te\textsubscript{3} as a function of magnetic field (\textit{H}) and temperature (\textit{T}). In Fig. \ref{fig:figure3}A, we present \textchi{ }vs. \textit{T} curves with \textit{H} along the SIP, LIP, and OOP morphological directions (see Fig. \ref{fig:figure1}A for reference to crystallographic axes). When cooling the sample from room temperature, the measurements show a paramagnetic-to-antiferromagnetic phase transition, characterized by a peak in \textchi{ }at the Néel temperature \textit{T}\textsubscript{N} = 164 $\pm$ 2 K (Fig. \ref{fig:figures3}, \ref{fig:figures4} for additional data). The Néel temperature is also confirmed through \textit{H}-dependent heat capacity measurements (Fig. \ref{fig:figures5}). Below the Néel temperature,  the zero-field-cooled (ZFC) and field-cooled (FC) branches separate, signaling the emergence of a spin-glassy phase below $\approx$100 K (Fig. \ref{fig:figures6}-\ref{fig:figures9} for a more detailed analysis)\cite{RN25}. This behavior, often associated with competing antiferromagnetic and ferromagnetic interactions, has been observed in other TaFe\textsubscript{1+\textit{y}}Te\textsubscript{3} stoichiometries\cite{RN25} and certain Fe\textsubscript{\textit{x}}Mn\textsubscript{1–\textit{x}}TiO\textsubscript{3} systems where long-range antiferromagnetic and re-entrant spin-glass phases can coexist\cite{RN28,RN29,RN30,RN31}. The extracted Weiss temperatures (\texttheta\textsubscript{w}) in the paramagnetic phase convey strong ferromagnetic coupling, with values of 154 K, 158 K, and 149 K for the OOP, LIP, and SIP directions, respectively (inset of Fig. \ref{fig:figure3}A). The Curie-Weiss fit estimates an effective magnetic moment (\textmu\textsubscript{eff}) ranging between 4.26-4.28 \textmu\textsubscript{B} for all morphological directions, in line with the expected local moment of 4 \textmu\textsubscript{B} for a majority high-spin tetrahedral Fe\textsuperscript{2+} ion (Fig. \ref{fig:figures23})\cite{RN32}.

Fig. \ref{fig:figure3}B-D plots the magnetization (\textit{M}) vs. applied \textit{H} along all morphological directions (see Fig. \ref{fig:figure1}A) up to 35 T at various \textit{T}. The OOP- and SIP-oriented \textit{M} vs. \textit{H} curves at 1.5 K reveal a series of steps in \textit{M} between 9-15 T after which \textit{M} continues to increase until saturating at 29 T and 30 T, respectively (Fig. \ref{fig:figure3}B, D, and Fig. \ref{fig:figures10}). The exact nature of these steps is unclear without further measurements of the high-field magnetic structure, but are likely due to metamagnetic transitions (see Fig. \ref{fig:figures11} and \ref{fig:figures12} for additional data).  The LIP-oriented \textit{M} vs. \textit{H} curve at 1.5 K is linear up to 27 T after which it saturates, indicating a gradual spin canting process (Fig. \ref{fig:figure3}C). This confirms the LIP direction as the hard axis (Fig. \ref{fig:figure3}C and Fig. \ref{fig:figures10}). At 1.5 K, the sample reaches a saturation magnetic moment for all axial orientations ranging between 2.05-2.12 \textmu\textsubscript{B}, which is significantly smaller than the \textmu\textsubscript{eff} determined for the high-\textit{T} paramagnetic phase. As expected, increasing \textit{T} decreases the saturation magnetic moment and the \textit{H} value at which the moment saturates. 

%%%%%%%%%%%%%%%%%%%%%%%%%%%%%%%%%%%%%%%%%%%%%%%%%%%%%%%%%%%%%%%%%%%%%%%%%%%%%%%%%%%%%%%%%%%%%%%%%%%%%%%%%%%%
%Figure 4
%%%%%%%%%%%%%%%%%%%%%%%%%%%%%%%%%%%%%%%%%%%%%%%%%%%%%%%%%%%%%%%%%%%%%%%%%%%%%%%%%%%%%%%%%%%%%%%%%%%%%%%%%%%%
\begin{figure*}[t]
\includegraphics[width=6.9in]{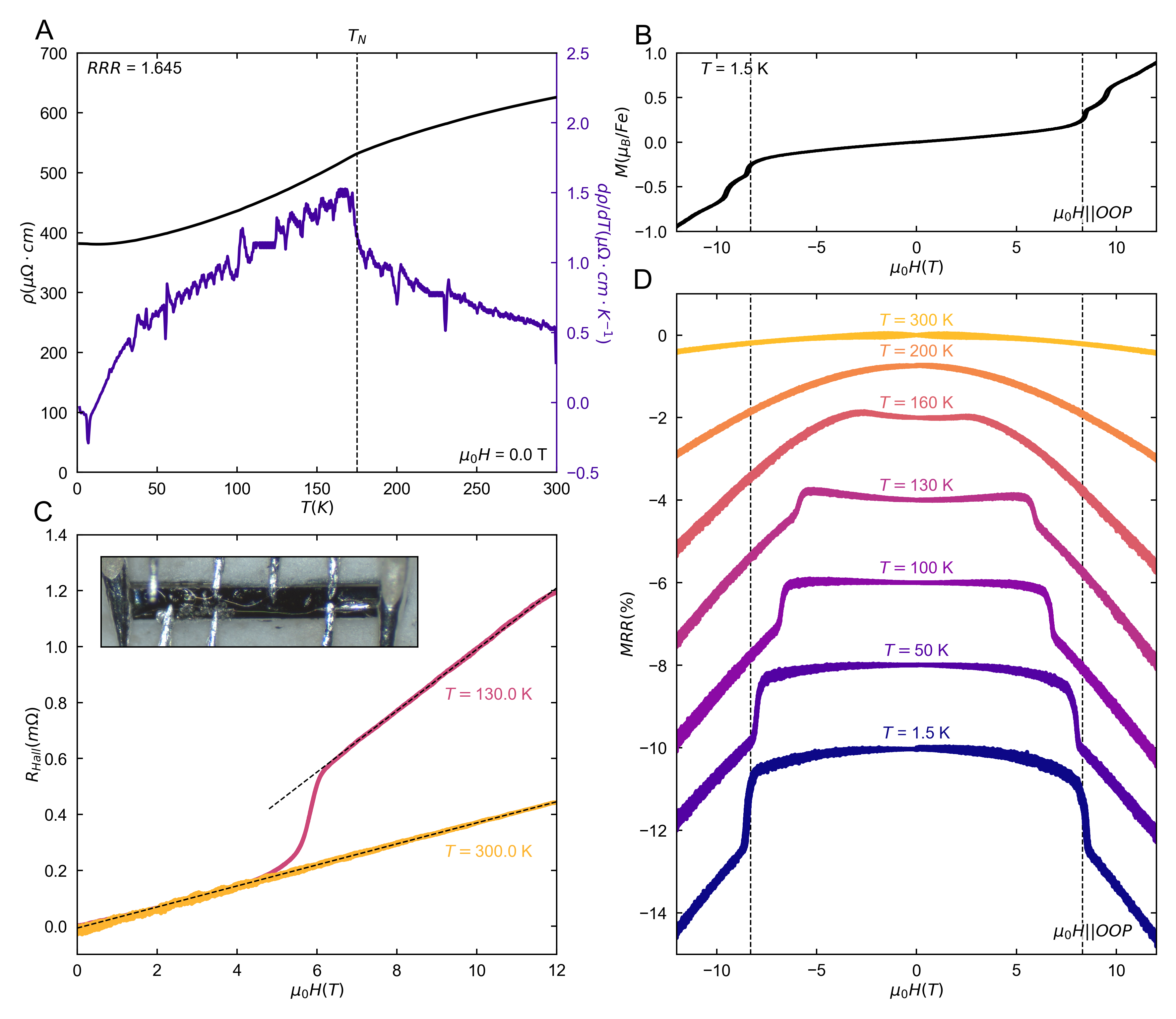}% Here is how to import EPS art
\centering
\caption[Electronic transport properties of TaFe\textsubscript{1.14}Te\textsubscript{3}]{\label{fig:figure4} Electronic transport properties of TaFe\textsubscript{1.14}Te\textsubscript{3}. A) \textrho{ }(left) and d\textrho/d\textit{T} (right) versus \textit{T} at zero \textit{H}. The extracted residual resistivity ratio (\textit{RRR} = \textit{R}(300K)/\textit{R}(2K)) is given in the inset. The extracted \textit{T}\textsubscript{N} is shown by a black dashed line. B) \textit{M} versus \textit{H} at 1.5 K for \textit{H} parallel to the OOP direction. C) \textit{R}\textsubscript{Hall} versus \textit{H} above (solid yellow line) and below (solid magenta line) \textit{T}\textsubscript{N}. Two distinct linear regions (denoted by dashed black lines) are observed below \textit{T}\textsubscript{N}, whereas a single linear region is observed above \textit{T}\textsubscript{N}. Additional traces at different temperatures can be found in Fig. \ref{fig:figures18}. An optical image of the device is given in the inset. D) Magnetoresistance ratio, \textit{MRR} = [(\textit{R}(\textit{H})-\textit{R}(\textit{H}=0))/\textit{R}(\textit{H}=0)]$\times$100, versus \textit{H} at various \textit{T} for fields parallel to the OOP direction. Both forward and backward traces are shown. The \textit{T} at which each trace was taken is given in the inset. The dashed black line denotes the 1.5 K metamagnetic transition. }
\end{figure*}
%%%%%%%%%%%%%%%%%%%%%%%%%%%%%%%%%%%%%%%%%%%%%%%%%%%%%%%%%%%%%%%%%%%%%%%%%%%%%%%%%%%%%%%%%%%%%%%%%%%%%%%%%%%%
%%%%%%%%%%%%%%%%%%%%%%%%%%%%%%%%%%%%%%%%%%%%%%%%%%%%%%%%%%%%%%%%%%%%%%%%%%%%%%%%%%%%%%%%%%%%%%%%%%%%%%%%%%%%

Our measured saturation moments of $\approx$2 \textmu\textsubscript{B} are consistent with those assigned previously based on neutron diffraction measurements of the antiferromagnetic phase\cite{RN23}. The discrepancy between \textmu\textsubscript{eff} and the measured saturation magnetic moment implies that both local moment and itinerant magnetism coexist in TaFe\textsubscript{1.14}Te\textsubscript{3}. For further insight, we performed first-principles DFT calculations on TaFe\textsubscript{1+\textit{y}}Te\textsubscript{3} with \textit{y} = 0, 0.25, and 1 with various magnetic orderings (see methods for details). We find that the antiferromagnetic order shown in Fig. \ref{fig:figure1}B, with intraplane ferromagnetic order and interplane antiferromagnetic order, is always electronically stable and typically lower in energy than other magnetic orderings (including non-magnetic), although the energy differences are often times small. Löwdin charge analysis of Fe atoms gives a positive charge of 0.3-0.5 and a magnetic moment of 2.2-2.4 \textmu\textsubscript{B}, for both Fe1 and Fe2 (Tab. \ref{tab:tables3}), in good agreement with our measured saturation moment. Because DFT with common functionals is not appropriate for strongly correlated electrons, the agreement with experiment supports the picture that electron itinerancy is responsible for the suppression of the magnetic moment. Although the character of the bands near the Fermi energy (\textit{E}\textsubscript{F}) is found to be a complex and nearly equal mixture of Fe 3d, Ta 5d and Te 5p orbitals (Fig. \ref{fig:figures20},\ref{fig:figures21}), all DFT calculations predict a metallic electronic structure, whose properties we study further by magnetotransport measurements.

In Fig. \ref{fig:figure4}, we report the magnetotransport of TaFe\textsubscript{1.14}Te\textsubscript{3} single crystals. Transport devices were fabricated using a micro-soldering technique which consists of drawing micron-scale contacts with Field’s metal using a micromanipulator (see methods for details)\cite{RN33andS2,RN34andS1}. Fig. \ref{fig:figure4}A plots the zero-field resistivity (\textrho) of TaFe\textsubscript{1.14}Te\textsubscript{3} versus \textit{T}. Overall metallic behavior is observed with \textrho{ }decreasing with decreasing \textit{T}. At \textit{T} = 175 $\pm$ 3 K, there is a sharp jump in the derivative of \textrho{ }versus \textit{T}, which is attributed to the onset of antiferromagnetic order. Below this temperature \textrho{ }varies approximately as \textit{T}\textsuperscript{2}, saturating to a constant value at low \textit{T}.

Fig. \ref{fig:figure4}D presents the \textit{H} dependence of the magnetoresistance ratio (\textit{MRR}). \textit{MRR} is defined as \textit{MRR} =  [(\textit{R}(\textit{H})-\textit{R}(\textit{H}=0))/\textit{R}(\textit{H}=0)]$\times$100, where \textit{H} is oriented along the OOP direction. Above \textit{T}\textsubscript{N}, the system is in the paramagnetic phase characterized by a broad negative \textit{MRR} (\textit{nMRR}) due to the field-induced suppression of spin-flip scattering between conduction electrons and local magnetic moments. Well below \textit{T}\textsubscript{N} at 1.5 K, we observe a negligible \textit{MRR} up to a critical field of \textit{H}\textsubscript{C} = 8.5 $\pm$ 0.2 T above which, the absolute value of \textit{MRR} sharply increases and continues to increase linearly for \textit{H}$>$\textit{H}\textsubscript{C}. The response of \textit{MRR} closely follows the measured \textit{M} vs \textit{H} traces (Fig. \ref{fig:figure4}B and Fig. \ref{fig:figures13}-\ref{fig:figures17}), indicating that \textit{H}\textsubscript{C} is correlated with the jump in \textit{M} vs \textit{H}. As \textit{T} increases, \textit{H}\textsubscript{C} decreases towards zero at \textit{T}\textsubscript{N}, consistent with the magnetization measurements. These results demonstrate that TaFe\textsubscript{1.14}Te\textsubscript{3} exhibits a relatively strong coupling between electrical properties and magnetic order. The fact that \textit{MRR} decreases as \textit{M} increases suggests the primary mechanism could be a field-induced suppression of interlayer spin-flip scattering. However, elucidating the exact mechanism behind this coupling will require further measurements of the field-induced magnetic structure. 

The coupling between electrical and magnetic properties can also be seen in the Hall measurements presented in Fig. \ref{fig:figure4}C. In magnetic materials, the Hall resistance (\textit{R}\textsubscript{Hall}) is typically determined by two components arising from the external and internal magnetic fields\cite{RN35}. In general, \textit{R}\textsubscript{Hall} is proportional to the applied \textit{H} (the slope of which is related to the carrier density) plus the sample magnetization. These two contributions can be seen clearly seen in Hall measurements across the magnetic transition (Fig. \ref{fig:figure4}C; see also Fig. \ref{fig:figures13}-\ref{fig:figures18}). Above \textit{T}\textsubscript{N} (at 300 K), since the sample has negligible magnetization (Fig. \ref{fig:figures15}), the \textit{R}\textsubscript{Hall} versus \textit{H} is linear, from which we extract a corresponding carrier density of 1.30 $\pm$ 0.03 $\times$ 10\textsuperscript{21} cm\textsuperscript{–3}, consistent with previous measurements on similar TaFe\textsubscript{1+\textit{y}}Te\textsubscript{3} stoichiometries\cite{RN24,RN26}. Below \textit{T}\textsubscript{N} (130 K), \textit{R}\textsubscript{Hall} versus \textit{H} is linear at low \textit{H} with the same slope as the 300 K trace, corresponding to the normal Hall contribution from external \textit{H}. Once \textit{H} reaches \textit{H}\textsubscript{C}, \textit{R}\textsubscript{Hall} sharply increases and then continues to increase linearly with \textit{H} for \textit{H}$>$\textit{H}\textsubscript{C}. The slope of \textit{R}\textsubscript{Hall} versus \textit{H} when \textit{H}$>$\textit{H}\textsubscript{C} is larger than the slope for \textit{H}$<$\textit{H}\textsubscript{C} due to the additional Hall component from the linear sample magnetization (Fig. \ref{fig:figures14}). The slope of \textit{R}\textsubscript{Hall} versus \textit{H} for both \textit{H}$>$\textit{H}\textsubscript{C} and \textit{H}$<$\textit{H}\textsubscript{C} has a non-monotonic temperature dependence which may arise from temperature-dependent magnetic fluctuations (Fig. \ref{fig:figures18})\cite{RN24}. It’s worth emphasizing that other factors could contribute to \textit{R}\textsubscript{Hall}, such as temperature- and field-dependent changes in the band structure. However, our results bear similarities to other TaFe\textsubscript{1+\textit{y}}Te\textsubscript{3} stoichiometries\cite{RN24,RN26} and ARPES measurements on TaFe\textsubscript{1.14}Te\textsubscript{3} (as will be seen below) show only minor changes in the band structure across the magnetic transition, indicating \textit{R}\textsubscript{Hall} is primarily determined by the sample carrier density and the sample magnetization (see Fig. \ref{fig:figures13}-\ref{fig:figures18} for additional data). This clear effect of the sample magnetization on the measured \textit{R}\textsubscript{Hall} further demonstrates the strong coupling between magnetism and electronic properties in TaFe\textsubscript{1.14}Te\textsubscript{3}. 

%%%%%%%%%%%%%%%%%%%%%%%%%%%%%%%%%%%%%%%%%%%%%%%%%%%%%%%%%%%%%%%%%%%%%%%%%%%%%%%%%%%%%%%%%%%%%%%%%%%%%%%%%%%%
%Figure 5
%%%%%%%%%%%%%%%%%%%%%%%%%%%%%%%%%%%%%%%%%%%%%%%%%%%%%%%%%%%%%%%%%%%%%%%%%%%%%%%%%%%%%%%%%%%%%%%%%%%%%%%%%%%%
\begin{figure*}[t]
\includegraphics[width=6.9in]{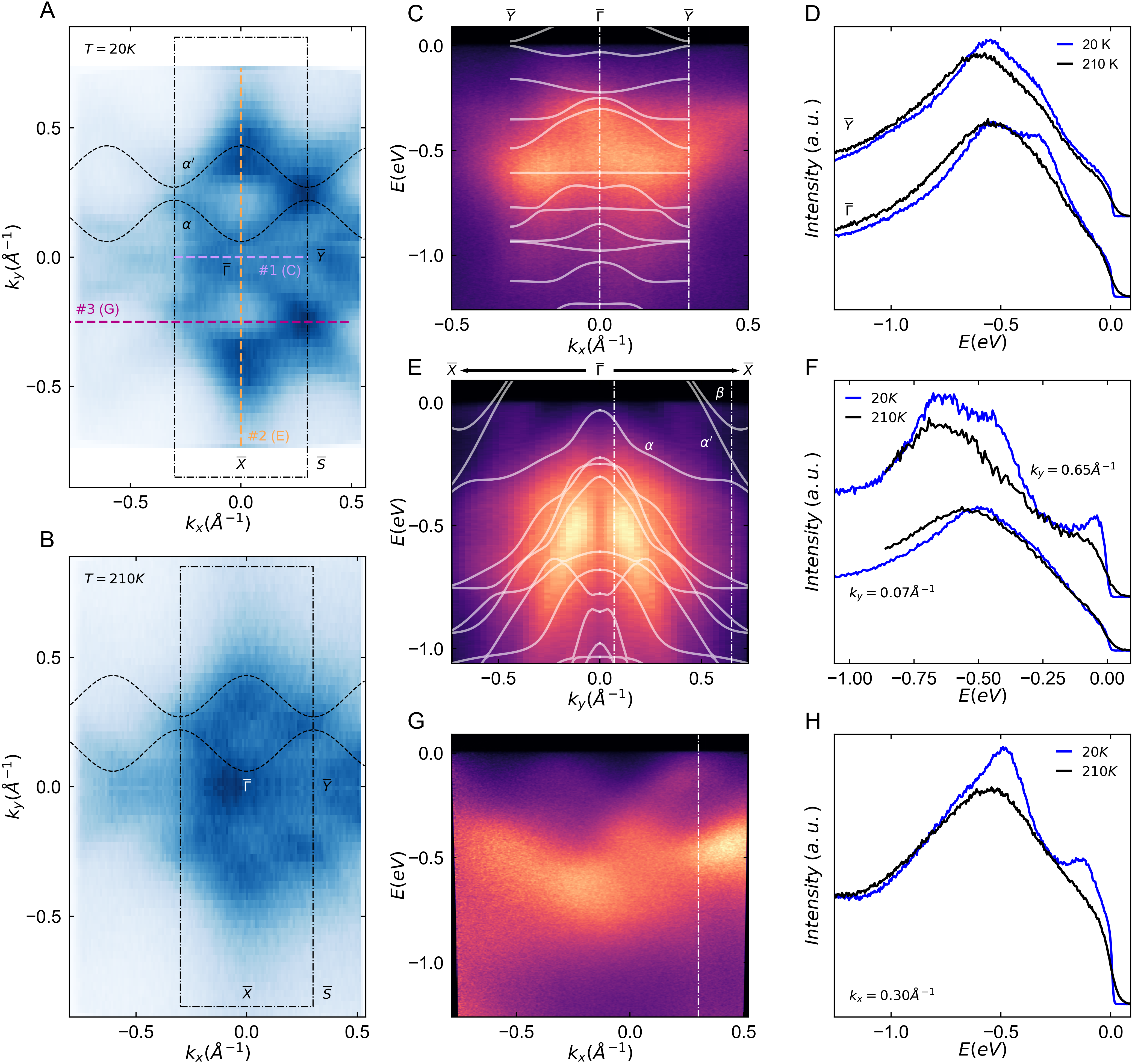}% Here is how to import EPS art
\centering
\caption[Evolution of the Fermi surface and electronic states of TaFe\textsubscript{1.14}Te\textsubscript{3} across the magnetic transition.]{\label{fig:figure5} Evolution of the Fermi surface and electronic states of TaFe\textsubscript{1.14}Te\textsubscript{3} across the magnetic transition. ARPES Fermi surface maps at 20 K (A) and 210 K (B). Dark blue (white) colors correspond to high (low) photoelectron intensity. The Fermi surfaces are symmetrized at \textit{k}\textsubscript{y} = 0 \AA\textsuperscript{-1}. In both (A) and (B), the black dashed rectangle represents the surface Brillouin zone. The $\bar{\Gamma}\bar{X}$ and $\bar{\Gamma}\bar{Y}$ directions are along (LIP) and perpendicular (SIP) to the Fe-Te chains, respectively. C) ARPES intensity map along the $\bar{Y}-\bar{\Gamma}-\bar{Y}$ direction is shown as cut \#1 (purple dashed line in (A)). Calculated non-magnetic band structure is overlayed as semi-transparent white lines.  D) Energy distribution curves (EDCs) at $\bar{\Gamma}$ and $\bar{Y}$ symmetry points (vertical dashed white lines in (C)) at 20 K (solid blue lines) and 210 K (solid black lines). E) ARPES intensity map along the $\bar{X}-\bar{\Gamma}-\bar{X}$ direction is shown as cut \#2 (orange dashed line in (A)). Calculated non-magnetic band structure is overlayed as semi-transparent white lines. F) EDCs at \textit{k}\textsubscript{y} = 0.07 \AA\textsuperscript{-1} (\textalpha{}) and \textit{k}\textsubscript{y} = 0.65 \AA\textsuperscript{-1} (\textbeta{}) (vertical dashed white lines in (E)) at 20 K (solid blue lines) and 210 K (solid black lines). G) ARPES intensity map along cut \#3 (magenta dashed lined in (A)). H) EDCs at \textit{k}\textsubscript{x} = 0.30 \AA\textsuperscript{-1}   (vertical dashed white line in (G)) at 20 K and 210 K. All data were acquired with a photon energy of 21.22 eV. For (C), (E), and (G), purple (yellow) colors correspond to low (high) photoelectron intensity. All calculated band structures were scaled down by 1.4 and shifted down by 0.2 eV.}
\end{figure*}
%%%%%%%%%%%%%%%%%%%%%%%%%%%%%%%%%%%%%%%%%%%%%%%%%%%%%%%%%%%%%%%%%%%%%%%%%%%%%%%%%%%%%%%%%%%%%%%%%%%%%%%%%%%%
%%%%%%%%%%%%%%%%%%%%%%%%%%%%%%%%%%%%%%%%%%%%%%%%%%%%%%%%%%%%%%%%%%%%%%%%%%%%%%%%%%%%%%%%%%%%%%%%%%%%%%%%%%%%

To experimentally probe the electronic structure of TaFe\textsubscript{1.14}Te\textsubscript{3}, we performed ARPES on a freshly cleaved surface (vdW plane) under ultra-high vacuum (UHV) across the magnetic transition (Fig. \ref{fig:figure5}). The Fermi surface at 20 and 210 K are shown in Fig. \ref{fig:figure5}A, B, respectively. Two sets of periodic wave-like Fermi surfaces (FSs) are identified, indicated by \textalpha{ }and \textalpha{}' in Fig. \ref{fig:figure5}A. The modulated FSs perpendicular to Fe1-Te chain directions (i.e., $\bar{\Gamma}\bar{Y}$) point to significant inter-chain interactions. The FS features look similar between the paramagnetic and antiferromagnetic phases, except for the thermal broadening, consistent with the previous report\cite{RN27}. The broadened spectroscopic features are likely a result of the disorder associated with the Fe2 site.

The ARPES intensity maps along the $\bar{Y}-\bar{\Gamma}-\bar{Y}$ and $\bar{X}-\bar{\Gamma}-\bar{X}$ paths in the low-\textit{T} antiferromagnetic phase are shown in Fig. \ref{fig:figure5}C, E (cut \#1 and \#2 in Fig. \ref{fig:figure5}A). ARPES intensity maps are compared to our DFT band structures of non-magnetic TaFeTe\textsubscript{3}, where the \textit{E}\textsubscript{F} was empirically shifted to best match the ARPES data (additional orbital-resolved band structures and partial density of states can be found in Fig. \ref{fig:figures20},\ref{fig:figures21}). Along the $\bar{X}-\bar{\Gamma}-\bar{X}$ path, we observe two hole-like bands, \textalpha{ }and \textalpha{}', crossing \textit{E}\textsubscript{F}. Another less dispersive band (\textbeta{}) is also observed near \textit{E}\textsubscript{F}. Overall, experimental ARPES intensity maps show reasonable agreement with our DFT calculations, when the bandwidth of the latter is reduced by a factor of 1.4 (Fig. \ref{fig:figure5}C, E). The necessity of this rescaling signals relatively strong electron correlations that are not captured by standard DFT. Fig. \ref{fig:figure5}D displays the energy distribution curves (EDCs) at $\bar{\Gamma}$ and $\bar{Y}$ points across the magnetic transition. We observe a distinct change in spectral intensity and a transfer of spectral weight from high-to-low binding energy, along with a slight shift in the energy of the bands. A more detailed \textit{T} dependence reveals that the most significant changes occur below $\approx$100 K (Fig. \ref{fig:figures22}). The evolution of EDCs across the magnetic transition for the \textalpha{ }and \textbeta{ }bands is shown in Fig. \ref{fig:figure5}F. Below the magnetic transition, the \textbeta{ }band shows an enhancement in quasiparticle (QP) spectral weight near \textit{E}\textsubscript{F}. However, no such change was observed for the \textalpha{ }band. ARPES intensity maps along cut \#3 (shown as a dashed line in Fig. \ref{fig:figure5}A) and EDCs are shown in Fig. \ref{fig:figure5}G, H, respectively. A strong enhancement of the integrated spectral weight of $\approx$16\% is observed within 0 to –1.25 eV in the antiferromagnetic phase, relative to the paramagnetic phase. These combined results reflect that the magnetic transition is driven by the electronic structure reconstruction rather than Fermi surface nesting. 

Our results suggest that the electronic states undergo a band-dependent reconstruction across the magnetic transition due to the interplay between orbital and spin degrees of freedom. This behavior is known to occur in other systems containing local moments and itinerant electrons\cite{RN36,RN37,RN38,RN39}, such as heavy fermion compounds\cite{RN40}, Fe-based superconductors\cite{RN41,RN42}, and other strongly-correlated transition-metal oxides\cite{RN43,RN44}. The coexistence of itinerant and local-moment magnetism in TaFe\textsubscript{1.14}Te\textsubscript{3}, combined with the fact that TaFe\textsubscript{1.14}Te\textsubscript{3} can be exfoliated down to atomically thin flakes, provides new opportunities to study such correlated magnetic systems in the 2D limit. 

%%%%%%%%%%%%%%%%%%%%%%%%%%%%%%%%%%%%%%%%%%%%%%%%%%%%%%%%%%%%%%%%%%%%%%%%%%%%%%%%%%%%%%%%%%%%%%%%%%%%%%%%%%%%
%Conclusion
%%%%%%%%%%%%%%%%%%%%%%%%%%%%%%%%%%%%%%%%%%%%%%%%%%%%%%%%%%%%%%%%%%%%%%%%%%%%%%%%%%%%%%%%%%%%%%%%%%%%%%%%%%%%
\section*{\label{sec:conclusions} Conclusions}
Through a combination of magnetic, electronic, and transport measurements, supported by first-principles DFT calculations, we resolve the interplay between the magnetic and electronic properties in the layered vdW magnet TaFe\textsubscript{1.14}Te\textsubscript{3}. We performed high-field magnetometry revealing the full direction-dependent magnetic properties. Calculations and magnetotransport measurements confirm the metallic nature of TaFe\textsubscript{1.14}Te\textsubscript{3} and display strong coupling between electronic transport properties and magnetic order. Temperature-dependent ARPES measurements show minor band shifts and strong spectral weight renormalization across the magnetic transition. This, combined with a low observed saturation moment of 2.05-2.12 \textmu\textsubscript{B} found in both the high-field magnetometry measurements and DFT calculations, indicate a complex interplay between local moment and itinerant magnetism in TaFe\textsubscript{1.14}Te\textsubscript{3}. Finally, we demonstrate the isolation of air-stable few-layer flakes of TaFe\textsubscript{1.14}Te\textsubscript{3} down to the monolayer limit. Altogether, our results establish TaFe\textsubscript{1.14}Te\textsubscript{3} as an exciting material for 2D spintronics, manifesting both robust A-type antiferromagnetism, metallic transport properties, and stability under ambient conditions. Furthermore, the intersection of local-moment and itinerant magnetism typically occurs in correlated electronic systems, offering TaFe\textsubscript{1.14}Te\textsubscript{3} as a promising platform for studying magnetism in 2D correlated electronic materials.

%%%%%%%%%%%%%%%%%%%%%%%%%%%%%%%%%%%%%%%%%%%%%%%%%%%%%%%%%%%%%%%%%%%%%%%%%%%%%%%%%%%%%%%%%%%%%%%%%%%%%%%%%%%%
%methods
%%%%%%%%%%%%%%%%%%%%%%%%%%%%%%%%%%%%%%%%%%%%%%%%%%%%%%%%%%%%%%%%%%%%%%%%%%%%%%%%%%%%%%%%%%%%%%%%%%%%%%%%%%%%
\section*{\label{sec:methods} Methods}
\subsection*{\label{subsec:synthesis} Synthesis of TaFe\textsubscript{1.14}Te\textsubscript{3}} 
Single crystals of TaFe\textsubscript{1.14}Te\textsubscript{3} were grown by chemical vapor transport. Tantalum, iron, tellurium, and TeCl\textsubscript{4} powders were combined in a 1:1.15:3:0.02 mole ratio, pressed into a pellet, and sealed in a fused silica tube (7 mm in diameter and approximately 40 cm in length) under vacuum. The sealed tube was loaded into a three-zone furnace and heated to a temperature gradient of 700$^\circ$C to 600$^\circ$C over 24 hours, with the pellet placed in the hot zone. After seven days, needle-like crystals were obtained in the sink end of the tube. 

\subsection*{\label{subsec:EDX} Energy Dispersive X-ray Spectroscopy (EDX) and Scanning Electron Microscopy (SEM)}
SEM images were acquired on a Zeiss SIGMA VP scanning electron microscope with a Bruker XFlash 6$|$30 Detector for EDX measurements. Single TaFe\textsubscript{1.14}Te\textsubscript{3} crystals were washed with hexanes, placed on carbon tape, and the surface was freshly cleaved with Scotch Magic tape immediately before loading into the chamber. SEM images were acquired with a beam energy of 5 kV. EDX spectra were acquired with a beam energy of 18 kV. The EDX collection time was automatically set by the Bruker ESPRIT 2 software using the “precise” acquisition setting. Elemental compositions and atomic percentages were estimated by integrating under the characteristic spectrum peaks for each element using Bruker ESPRIT 2 software.

\subsection*{\label{subsec:VSM} Vibrating Sample Magnetometry (VSM) up to 9T}
Vibrating sample magnetometry was conducted on a Quantum Design PPMS Dynacool system. TaFe\textsubscript{1.14}Te\textsubscript{3} single crystals were attached to a quartz paddle using GE varnish (which was cured at room temperature under ambient conditions for 45 minutes) and oriented with long in-plane (LIP), short in-plane (SIP), or out-of-plane (OOP) direction parallel to the applied magnetic field. The same crystal was used for all direction-dependent measurements. Variable temperature magnetic susceptibility and magnetic field-dependent magnetization curves for each orientation were measured during the same measurement cycle. Samples were removed using a 1:1 ethanol/toluene solution, dried in air, and then remounted to the quartz paddle using GE varnish.

\subsection*{\label{subsec:VSMHF} Vibrating Sample Magnetometry (VSM) up to 35T}
The high-field magnetization data were obtained at the National High Magnetic Field Laboratory (NHMFL) using a conventional Vibrating Sample Magnetometer.  The measurement was performed on a few single crystal samples with a total mass of 19.6 mg glued on a Torlon (polyamidimide) sample holder with GE7031 varnish. A water-cooled resistive magnet was used to apply magnetic fields up to 35 T.

\subsection*{\label{subsec:STM} Scanning Tunneling Microscopy/Spectroscopy (STM/S)}
STM and STS measurements were conducted in a home-built STM under ultra-high vacuum at 7.5 K. The tungsten tunneling tip was electrochemically etched and calibrated against the Au(111) surface state prior to each sample approach. Crystals were cleaved in vacuum prior to measurement. Topography was acquired in constant current mode with a set bias of 400 mV and a set current of 50 pA. 

\subsection*{\label{subsec:TEM} Transmission Electron Microscopy (TEM)}
TaFe\textsubscript{1.14}Te\textsubscript{3} flakes were isolated through mechanical exfoliation of single crystals and placed on a TEM grid. TEM images were taken with a Thermofisher Scientific Talos F200X. The sample was slightly tilted to the desired orientation and bright-field mode imaging was set by using the Objective Aperture 100 μm.

\subsection*{\label{subsec:device} Electrical Transport Device Fabrication}
TaFe\textsubscript{1.14}Te\textsubscript{3} crystals were bonded to a 16-pin DIP socket using low-temperature non-conducting epoxy (Loctite EA 1C) and the top surface was freshly cleaved.  All contacts to bulk TaFe\textsubscript{1.14}Te\textsubscript{3} were made with Field’s metal using a microsoldering technique\cite{RN34andS1,RN33andS2}. A micromanipulator and heated sample holder were placed under an optical microscope. The temperature of the sample holder was raised to 70$^\circ$C to melt the Field’s metal. The tip of the micromanipulator was lowered into the Field’s metal and slowly drawn out to produce a contact with a few-micron-wide tip. The contact was then aligned on top of the sample with the micromanipulator and placed on the sample while heating to 100$^\circ$C. Direct electrical connections from the Field’s metal contact to the dip socket gold pads were made by hand with silver paint (Dupont 4929N) and 25 {\textmu}m diameter aluminum wire-bonding wire.

\subsection*{\label{subsec:transport} Electrical Transport Measurements}
Longitudinal and Hall resistances were measured in a 4-terminal configuration using a Keithley 6221 as a high-precision current source and a Keithley 2182A as a DC voltmeter. The two instruments were operated in delta mode to reduce background noise. A source current of 5 mA was used for all measurements unless otherwise stated. Due to the morphology of the TaFe\textsubscript{1.14}Te\textsubscript{3} crystals, current and longitudinal resistance were measured parallel to the long in-plane direction (LIP) and the Hall voltage was measured along the short in-plane direction (SIP). Variable temperatures between 1.5 K and 300 K and magnetic fields between -12 T and 12 T were achieved in an Oxford TeslatronPT dry \textsuperscript{4}He cryostat.

\subsection*{\label{subsec:exfoliation} Exfoliation}
Thin flakes of TaFe\textsubscript{1.14}Te\textsubscript{3} were isolated using a two-step exfoliation procedure. First, single TaFe\textsubscript{1.14}Te\textsubscript{3} crystals were cleaved multiple times with Nitto REVALPHA thermal release tape\cite{RNS3}. Then, Scotch Magic tape was used to exfoliate the crystals from the thermal release tape and transfer them onto commercially available polydimethylsiloxane (PDMS) stamps. Once on the PDMS stamps, the thickness of exfoliated flakes was inferred from atomic force microscopy and optical contrast.

\subsection*{\label{subsec:ATM} Atomic Force Microscopy (AFM)}
Atomic force microscopy was performed in a Bruker Dimension Icon using OTESPA-R3 tips in tapping mode. Flake thicknesses were extracted using Gwyddion to measure the height difference between the substrate and the desired TaFe\textsubscript{1.14}Te\textsubscript{3} flake. 

\subsection*{\label{subsec:XPS} X-ray Photoemission Spectroscopy (XPS)}
XPS measurements were performed by a VersaProbe II (Physical Electronics) equipped with Al K\textalpha{ }X-rays and dual beam neutralization (10 eV Ar\textsuperscript{+} ions and electrons) at 45-degree measuring angle. The pass energies for survey and individual elements are 117.4 eV and 23.5 eV, respectively.

\subsection*{\label{subsec:SCXRD} Single crystal X-ray diffraction}
Single crystal diffraction measurements were collected on TaFe\textsubscript{1.14}Te\textsubscript{3} crystals using an Agilent Supernova single-crystal diffractometer. The crystals were mounted onto a MiTeGen MicroLoops holder with paratone oil.  The X-ray source was a Mo K\textalpha{ }micro-focus tube energized to 50 kV and 0.8 mA. The collection temperature was maintained between 300 K and 100 K using an Oxford instruments nitrogen cryostream. The data collection, integration, and reduction were performed using the Crysalis Pro software suite. The crystal structure was solved and refined using ShelXT and ShelXL respectively.

\subsection*{\label{subsec:PXRD} Powder X-ray diffraction (PXRD)}
Powder diffraction patterns were collected on a Panalytical Aeris diffractometer with a Cu K\textalpha{ }X-ray source energized to 40 kV and 15 mA. The X-ray beam was filtered with a Ni\textbeta{ }filter. The powder sample of TaFe\textsubscript{1.14}Te\textsubscript{3} (prepared by grinding single crystals in liquid nitrogen) was mounted on a Si-zero background holder which was spun to reduce preferred orientation. 

\subsection*{\label{subsec:heatcap} Heat capacity}
Heat capacity measurements were performed with a Quantum Design PPMS Dynacool system. The background heat capacity of the sample puck plus grease (Apiezon N grease) was measured over the desired temperature and magnetic field ranges before the sample was loaded onto the platform. The background was then subtracted from the total measured heat capacity (sample plus grease and puck) to obtain the heat capacity of the sample. 

\subsection*{\label{subsec:ARPES} Angle-resolved Photoemission Spectroscopy (ARPES)}
The sample was attached to a sample holder using Ag-epoxy and cleaved in-situ using Kapton tape, just before the ARPES measurements. The ARPES experiments were carried out at OASIS-laboratory at Brookhaven National Laboratory using a Scienta SES-R4000 electron spectrometer with monochromatized He-I\textalpha{ }(21.22 eV) radiation (VUV-5k)\cite{RNS4}. The total instrumental energy resolution was $\approx$10 meV. The angular resolution was better than 0.15\texttheta{ }and 0.4\texttheta{ }along and perpendicular to the slit of the analyzer, respectively. 

\subsection*{\label{subsec:powder} Preparation of TaFe\textsubscript{1.14}Te\textsubscript{3} powder samples  for magnetometry measurements under pressure}
Single crystals of TaFe\textsubscript{1.14}Te\textsubscript{3} were placed in a thin porcelain crucible with enough liquid N\textsubscript{2} to fully submerge the crystals. The crystals were then ground with a thermally equilibrated pestle for 5 mins. The resulting powder was rinsed with acetone to remove residual moisture from condensation.

\subsection*{\label{subsec:pressure} Determination of applied hydrostatic pressure for magnetometry measurements under pressure}

Since the superconducting (SC) critical temperature (\textit{T}\textsubscript{C}) of Pb is well-known to linearly depend upon applied hydrostatic pressure (\textit{P}) at a rate of d\textit{T}\textsubscript{C}/d\textit{P} = 0.379 K$\cdot$GPa\textsuperscript{-1}\cite{RNS5}, we use the measured \textit{T}\textsubscript{C} of Pb to determine the \textit{P} on TaFe\textsubscript{1.14}Te\textsubscript{3}. First, the Pb plus TaFe\textsubscript{1.14}Te\textsubscript{3} sample is zero-field cooled below the SC transition to 6 K. Then, the magnetic susceptibility (\textchi) versus temperature (\textit{T}) is measured with a small measuring field of 5 Oe (much smaller than the zero-temperature upper critical field of Pb\cite{RNS6}). \textchi{ }versus \textit{T} is measured at a rate of 0.05 K$\cdot$min\textsuperscript{-1} to ensure the transition is precisely resolved and traces with increasing and decreasing \textit{T} were measured. The Pb \textit{T}\textsubscript{C} is extracted by finding the condition where \textchi{ }= 0.5\textchi\textsubscript{N} (\textchi\textsubscript{N} is the susceptibility in the normal state).

\subsection*{\label{subsec:VSMpressure} Vibrating sample magnetometry under hydrostatic pressure}
All vibrating sample magnetometry under \textit{P} was conducted on a Quantum Design PPMS Dynacool system using the commercially-available HMD high-pressure cell. Before and after the VSM measurements, PXRD was used to confirm there was no significant change in structure upon grinding or after applying maximum pressure (Fig. \ref{fig:figures9}). The powder was then combined with Daphne 7373 oil and a $\approx$1-2 mm long wire of Pb in a Teflon capsule, which was then inserted into the pressure cell. The variable temperature scans and field-dependent magnetic susceptibility curves for each \textit{P} were measured during the same measurement cycle. The measurements performed at different pressures were done sequentially with increasing pressure (from zero applied pressure up to the maximum achievable pressure). After the final maximum pressure measurement, the capsule containing the TaFe\textsubscript{1.14}Te\textsubscript{3} powder, Daphne 7373 oil, and the Pb manometer was removed, fixed to a brass paddle with GE varnish, and re-measured as a consistency check of the zero-pressure measurement.

\subsection*{\label{subsec:theory} Computational Methods}
The computational results were obtained using density functional theory, as implemented in the QUANTUM ESPRESSO package\cite{RNS7,RNS8}. We used projector-augmented wave pseudopotentials (Ta.pbe-spn-kjpaw\_psl.1.0.0.UPF, Fe.pbe-spn-kjpaw\_psl.1.0.0.UPF, and Te.pbe-n-kjpaw\_psl.1.0.0.UPF from http://www.quantum-espresso.org) with the Perdew-Burke-Ernzerhof exchange-correlation functional\cite{RNS9}, with nonlinear core correction and scalar relativistic effects. Gaussian smearing was set to 0.005 Ry. The wave function and density energy cutoffs were 100 Ry and 400 Ry, respectively. The Brillouin zone was sampled with a 7x7x7 \textit{k}-space mesh for the paramagnetic, primitive cells and with a 3x7x7 \textit{k}-space mesh for the 2x1x1 AFM supercells. The unit cell was taken from MaterialsProject (mp-8848)\cite{RNS10}. Relaxed positions of atoms within the unit cell were used.

%%%%%%%%%%%%%%%%%%%%%%%%%%%%%%%%%%%%%%%%%%%%%%%%%%%%%%%%%%%%%%%%%%%%%%%%%%%%%%%%%%%%%%%%%%%%%%%%%%%%%%%%%%%%
%acknowledgements
%%%%%%%%%%%%%%%%%%%%%%%%%%%%%%%%%%%%%%%%%%%%%%%%%%%%%%%%%%%%%%%%%%%%%%%%%%%%%%%%%%%%%%%%%%%%%%%%%%%%%%%%%%%%
\section*{\label{sec:ack} Acknowledgements}
We would like to thank Daniel Chica and Amymarie Bartholomew for their help in analyzing SCXRD data and Mike Ziebel for useful discussions on magnetic data. We would also like to acknowledge the use of Python-based data acquisition software developed by Maëlle Kapfer for all transport measurements. Synthesis and structural/compositional characterization of layered metallic antiferromagnets was supported by the National Science Foundation (NSF) through the Columbia University Materials Research Science and Engineering Center (MRSEC) on Precision-Assembled Quantum Materials DMR-2011738. Magnetotransport measurements were supported as part of Programmable Quantum Materials, an Energy Frontier Research Center funded by the U.S. Department of Energy (DOE), Office of Science, Basic Energy Sciences (BES), under award DE-SC0019443. Magnetic characterization was supported by the Air Force Office of Scientific Research award FA9550-22-1-0389. Imaging experiments were supported by the NSF CAREER award DMR-1751949. S.J.B. acknowledges support from the NSF Award DGE-2036197. The authors acknowledge the use of facilities and instrumentation supported by the NSF through Columbia University, Columbia Nano Initiative, and the Materials Research Science and Engineering Center DMR-2011738. The National High Magnetic Field Laboratory is supported by the NSF through NSF/DMR-1644779 and the State of Florida. The research at Brookhaven National Laboratory was supported by the US Department of Energy, Office of Basic Energy Sciences, Contract No. DESC0012704. 

%%%%%%%%%%%%%%%%%%%%%%%%%%%%%%%%%%%%%%%%%%%%%%%%%%%%%%%%%%%%%%%%%%%%%%%%%%%%%%%%%%%%%%%%%%%%%%%%%%%%%%%%%%%%
%Author contributions
%%%%%%%%%%%%%%%%%%%%%%%%%%%%%%%%%%%%%%%%%%%%%%%%%%%%%%%%%%%%%%%%%%%%%%%%%%%%%%%%%%%%%%%%%%%%%%%%%%%%%%%%%%%%
\section*{\label{sec:author} Author Contributions}
S.Y.H. and E.J.T. collected and analyzed the SCXRD data. S.Y.H. and R.W. optimized the synthesis procedure and synthesized the single crystals. S.Y.H. and E.J.T. performed the SEM/EDX and analyzed the data. S.T. performed the STM and STS experiments. A.Z. collected and analyzed the TEM data. S.J.B. performed DFT calculations. S.Y.H. and E.J.T. performed the low-field magnetometry. S.Y.H., E.J.T., and E-S.C. performed the high-magnetometry. S.Y.H. prepared the single-crystal transport devices and E.J.T. performed the transport measurements and analyzed the data. A.K. collected and analyzed the ARPES data. T-D.L. performed the XPS experiments. S.Y.H. and E.J.T. exfoliated few-layer flakes and performed the atomic force microscopy and contrast analysis. S.Y.H. and E.J.T. performed heat capacity measurements and analyzed the data. S.Y.H. performed and analyzed the PXRD data. The manuscript was prepared with input from all co-authors.

%%%%%%%%%%%%%%%%%%%%%%%%%%%%%%%%%%%%%%%%%%%%%%%%%%%%%%%%%%%%%%%%%%%%%%%%%%%%%%%%%%%%%%%%%%%%%%%%%%%%%%%%%%%%
%Supporting information
%%%%%%%%%%%%%%%%%%%%%%%%%%%%%%%%%%%%%%%%%%%%%%%%%%%%%%%%%%%%%%%%%%%%%%%%%%%%%%%%%%%%%%%%%%%%%%%%%%%%%%%%%%%%
\section*{\label{sec:SI} Supporting Information}
Supporting information is available at the end of this document after the references.

%%%%%%%%%%%%%%%%%%%%%%%%%%%%%%%%%%%%%%%%%%%%%%%%%%%%%%%%%%%%%%%%%%%%%%%%%%%%%%%%%%%%%%%%%%%%%%%%%%%%%%%%%%%%
%Conflicts of interest
%%%%%%%%%%%%%%%%%%%%%%%%%%%%%%%%%%%%%%%%%%%%%%%%%%%%%%%%%%%%%%%%%%%%%%%%%%%%%%%%%%%%%%%%%%%%%%%%%%%%%%%%%%%%
\section*{\label{sec:conflict} Conflict of Interest}
The authors declare no conflict of interest.

%%%%%%%%%%%%%%%%%%%%%%%%%%%%%%%%%%%%%%%%%%%%%%%%%%%%%%%%%%%%%%%%%%%%%%%%%%%%%%%%%%%%%%%%%%%%%%%%%%%%%%%%%%%%
%References
%%%%%%%%%%%%%%%%%%%%%%%%%%%%%%%%%%%%%%%%%%%%%%%%%%%%%%%%%%%%%%%%%%%%%%%%%%%%%%%%%%%%%%%%%%%%%%%%%%%%%%%%%%%%
\bibliography{TaFeTe_ref.bib}% Produces the bibliography via BibTeX.
\clearpage
%%%%%%%%%%%%%%%%%%%%%%%%%%%%%%%%%%%%%%%%%%%%%%%%%%%%%%%%%%%%%%%%%%%%%%%%%%%%%%%%%%%%%%%%%%%%%%%%%%%%%%%%%%%%
%Supporting information
%%%%%%%%%%%%%%%%%%%%%%%%%%%%%%%%%%%%%%%%%%%%%%%%%%%%%%%%%%%%%%%%%%%%%%%%%%%%%%%%%%%%%%%%%%%%%%%%%%%%%%%%%%%%
\onecolumngrid
\section*{\label{sec:SI_details} Supporting Information}
%\maketitle
\listoffigures
\listoftables
%redefine the figure labelling
\renewcommand\thefigure{S\arabic{figure}}    
\setcounter{figure}{0} 
%%%%%%%%%%%%%%%%%%%%%%%%%%%%%%%%%%%%%%%%%%%%%%%%%%%%%%%%%%%%%%%%%%%%%%%%%%%%%%%%%%%%%%%%%%%%%%%%%%%%%%%%%%%%
%Figure s1
%%%%%%%%%%%%%%%%%%%%%%%%%%%%%%%%%%%%%%%%%%%%%%%%%%%%%%%%%%%%%%%%%%%%%%%%%%%%%%%%%%%%%%%%%%%%%%%%%%%%%%%%%%%%
\begin{figure*}
\includegraphics[width=6.9in]{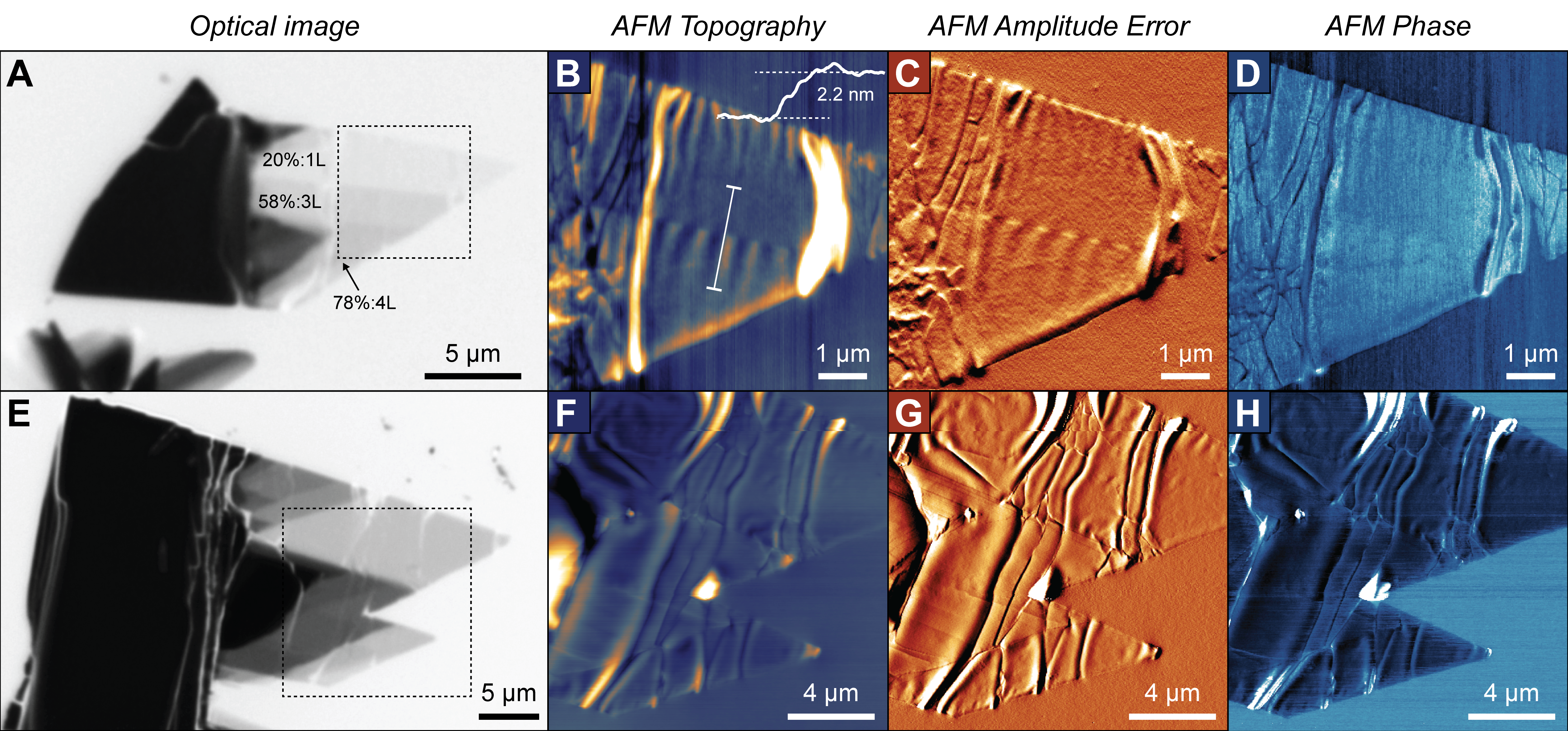}% Here is how to import EPS art
\centering
\caption[Mechanical exfoliation of TaFe\textsubscript{1.14}Te\textsubscript{3} flakes]{\label{fig:figures1} Mechanical exfoliation of TaFe\textsubscript{1.14}Te\textsubscript{3} flakes. A, E) Grey-scale optical images of two sample TaFe\textsubscript{1.14}Te\textsubscript{3} flakes exfoliated onto PDMS using the previously-described tape method. White and black colors correspond to the PDMS substrate and thick TaFe\textsubscript{1.14}Te\textsubscript{3} flakes, respectively. In (A), the extracted optical contrast for 1, 3, and 4L regions is denoted. B-D) tapping-mode atomic force microscope topography (B), amplitude error (C), and phase (D) for the region in (A) defined by the dashed black box. The inset of (B) shows a height-profile line cut of the step edge. F-H) tapping-mode atomic force microscope topography (F), amplitude error (G), and phase (H) for the region in (E) defined by the dashed black box.}
\end{figure*}
%%%%%%%%%%%%%%%%%%%%%%%%%%%%%%%%%%%%%%%%%%%%%%%%%%%%%%%%%%%%%%%%%%%%%%%%%%%%%%%%%%%%%%%%%%%%%%%%%%%%%%%%%%%%
%%%%%%%%%%%%%%%%%%%%%%%%%%%%%%%%%%%%%%%%%%%%%%%%%%%%%%%%%%%%%%%%%%%%%%%%%%%%%%%%%%%%%%%%%%%%%%%%%%%%%%%%%%%%

%%%%%%%%%%%%%%%%%%%%%%%%%%%%%%%%%%%%%%%%%%%%%%%%%%%%%%%%%%%%%%%%%%%%%%%%%%%%%%%%%%%%%%%%%%%%%%%%%%%%%%%%%%%%
%Figure s2
%%%%%%%%%%%%%%%%%%%%%%%%%%%%%%%%%%%%%%%%%%%%%%%%%%%%%%%%%%%%%%%%%%%%%%%%%%%%%%%%%%%%%%%%%%%%%%%%%%%%%%%%%%%%
\begin{figure*}[b]
\includegraphics[width=6.9in]{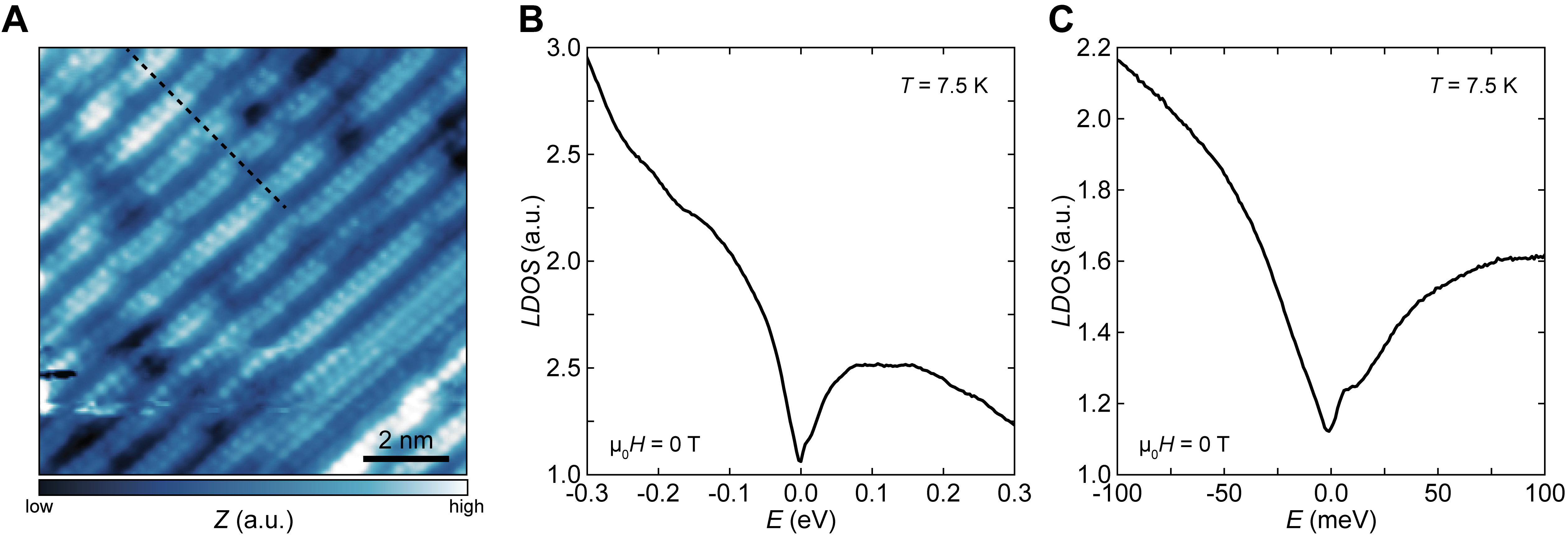}% Here is how to import EPS art
\centering
\caption[Scanning tunneling microscopy and spectroscopy of TaFe\textsubscript{1.14}Te\textsubscript{3}]{\label{fig:figures2} Scanning tunneling microscopy and spectroscopy of TaFe\textsubscript{1.14}Te\textsubscript{3}. A) Scanning tunneling microscopy image of bulk TaFe\textsubscript{1.14}Te\textsubscript{3} along the out-of-plane direction at \textit{T} = 7.5 K. The image was obtained in constant current mode (\textit{V}\textsubscript{b} = 400 mV, \textit{I} = 50 pA). B) Local density of states (LDOS) versus energy at 7.5 K and zero magnetic field. The trace is a line-average scan across multiple TaFe\textsubscript{1.14}Te\textsubscript{3} chains. The direction and length of the line-scan average are denoted by a black dashed line in (A). C) Zoom-in of the line-averaged LDOS versus energy trace at 7.5 K and zero magnetic field.}
\end{figure*}
%%%%%%%%%%%%%%%%%%%%%%%%%%%%%%%%%%%%%%%%%%%%%%%%%%%%%%%%%%%%%%%%%%%%%%%%%%%%%%%%%%%%%%%%%%%%%%%%%%%%%%%%%%%%
%%%%%%%%%%%%%%%%%%%%%%%%%%%%%%%%%%%%%%%%%%%%%%%%%%%%%%%%%%%%%%%%%%%%%%%%%%%%%%%%%%%%%%%%%%%%%%%%%%%%%%%%%%%%

%%%%%%%%%%%%%%%%%%%%%%%%%%%%%%%%%%%%%%%%%%%%%%%%%%%%%%%%%%%%%%%%%%%%%%%%%%%%%%%%%%%%%%%%%%%%%%%%%%%%%%%%%%%%
%Figure s3
%%%%%%%%%%%%%%%%%%%%%%%%%%%%%%%%%%%%%%%%%%%%%%%%%%%%%%%%%%%%%%%%%%%%%%%%%%%%%%%%%%%%%%%%%%%%%%%%%%%%%%%%%%%%
\begin{figure*}[t]
\includegraphics[width=6.9in]{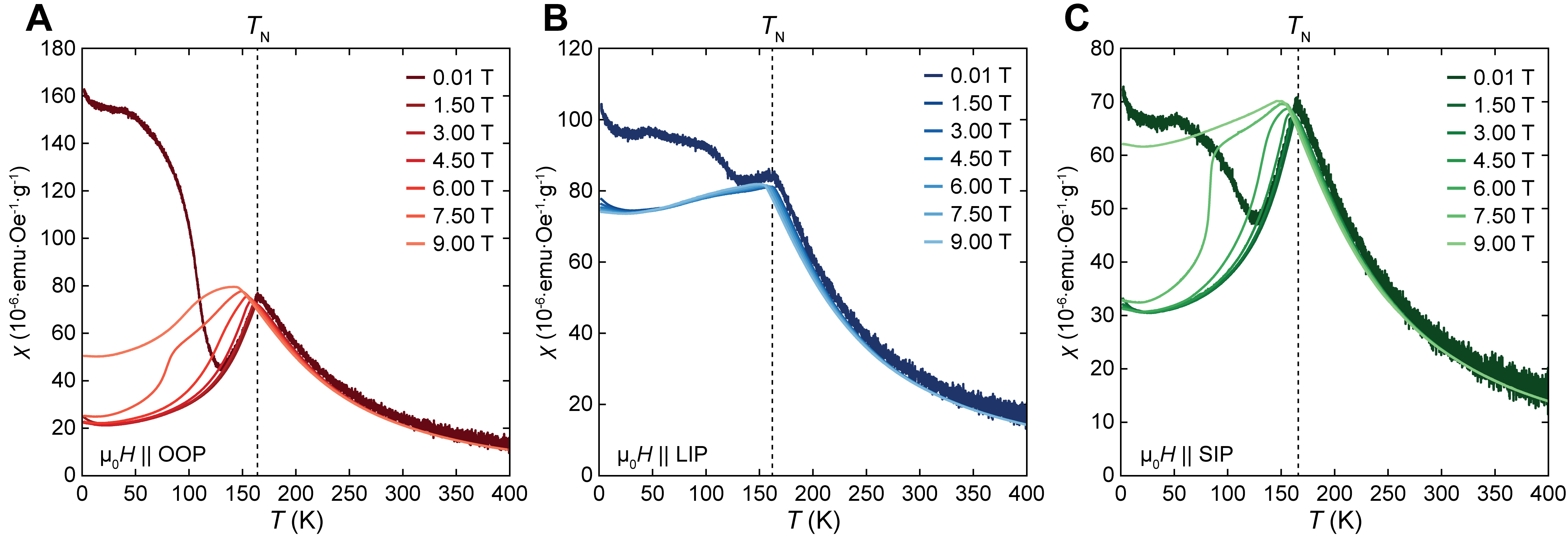}% Here is how to import EPS art
\centering
\caption[Field-dependent magnetic susceptibility on TaFe\textsubscript{1.14}Te\textsubscript{3}]{\label{fig:figures3} Field-dependent magnetic susceptibility on TaFe\textsubscript{1.14}Te\textsubscript{3}. A-C) Field-cooled magnetic susceptibility versus temperature for magnetic fields oriented along the out-of-plane (A), long in-plane (B), and short-in-plane (C) directions. The magnetic field at which each trace was taken is given in the insets. \textit{T}\textsubscript{N} is denoted by a dashed black line in each plot.}
\end{figure*}
%%%%%%%%%%%%%%%%%%%%%%%%%%%%%%%%%%%%%%%%%%%%%%%%%%%%%%%%%%%%%%%%%%%%%%%%%%%%%%%%%%%%%%%%%%%%%%%%%%%%%%%%%%%%
%%%%%%%%%%%%%%%%%%%%%%%%%%%%%%%%%%%%%%%%%%%%%%%%%%%%%%%%%%%%%%%%%%%%%%%%%%%%%%%%%%%%%%%%%%%%%%%%%%%%%%%%%%%%

%%%%%%%%%%%%%%%%%%%%%%%%%%%%%%%%%%%%%%%%%%%%%%%%%%%%%%%%%%%%%%%%%%%%%%%%%%%%%%%%%%%%%%%%%%%%%%%%%%%%%%%%%%%%
%Figure s4
%%%%%%%%%%%%%%%%%%%%%%%%%%%%%%%%%%%%%%%%%%%%%%%%%%%%%%%%%%%%%%%%%%%%%%%%%%%%%%%%%%%%%%%%%%%%%%%%%%%%%%%%%%%%
\begin{figure*}[b]
\includegraphics[width=6.9in]{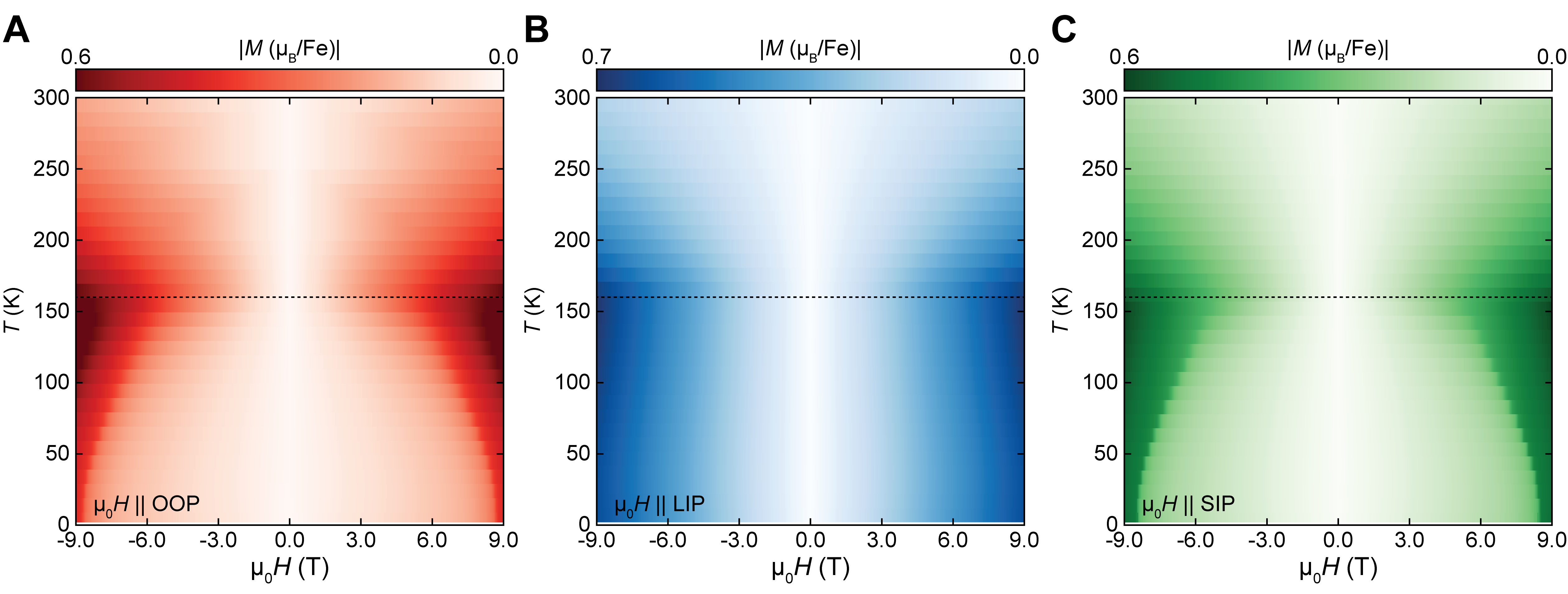}% Here is how to import EPS art
\centering
\caption[Temperature-dependent magnetization for TaFe\textsubscript{1.14}Te\textsubscript{3}]{\label{fig:figures4} Temperature-dependent magnetization data TaFe\textsubscript{1.14}Te\textsubscript{3}. A-C) Color maps of magnetization versus magnetic field (x-axis) and temperature (y-axis) for magnetic fields oriented along the out-of-plane (A), long in-plane (B), and short-in-plane (C) directions. The color bar scale is given above each plot. \textit{T}\textsubscript{N} is denoted by a dashed black line in each plot.}
\end{figure*}
%%%%%%%%%%%%%%%%%%%%%%%%%%%%%%%%%%%%%%%%%%%%%%%%%%%%%%%%%%%%%%%%%%%%%%%%%%%%%%%%%%%%%%%%%%%%%%%%%%%%%%%%%%%%
%%%%%%%%%%%%%%%%%%%%%%%%%%%%%%%%%%%%%%%%%%%%%%%%%%%%%%%%%%%%%%%%%%%%%%%%%%%%%%%%%%%%%%%%%%%%%%%%%%%%%%%%%%%%

%%%%%%%%%%%%%%%%%%%%%%%%%%%%%%%%%%%%%%%%%%%%%%%%%%%%%%%%%%%%%%%%%%%%%%%%%%%%%%%%%%%%%%%%%%%%%%%%%%%%%%%%%%%%
%Figure s5
%%%%%%%%%%%%%%%%%%%%%%%%%%%%%%%%%%%%%%%%%%%%%%%%%%%%%%%%%%%%%%%%%%%%%%%%%%%%%%%%%%%%%%%%%%%%%%%%%%%%%%%%%%%%
\begin{figure*}[t]
\includegraphics[width=6.9in]{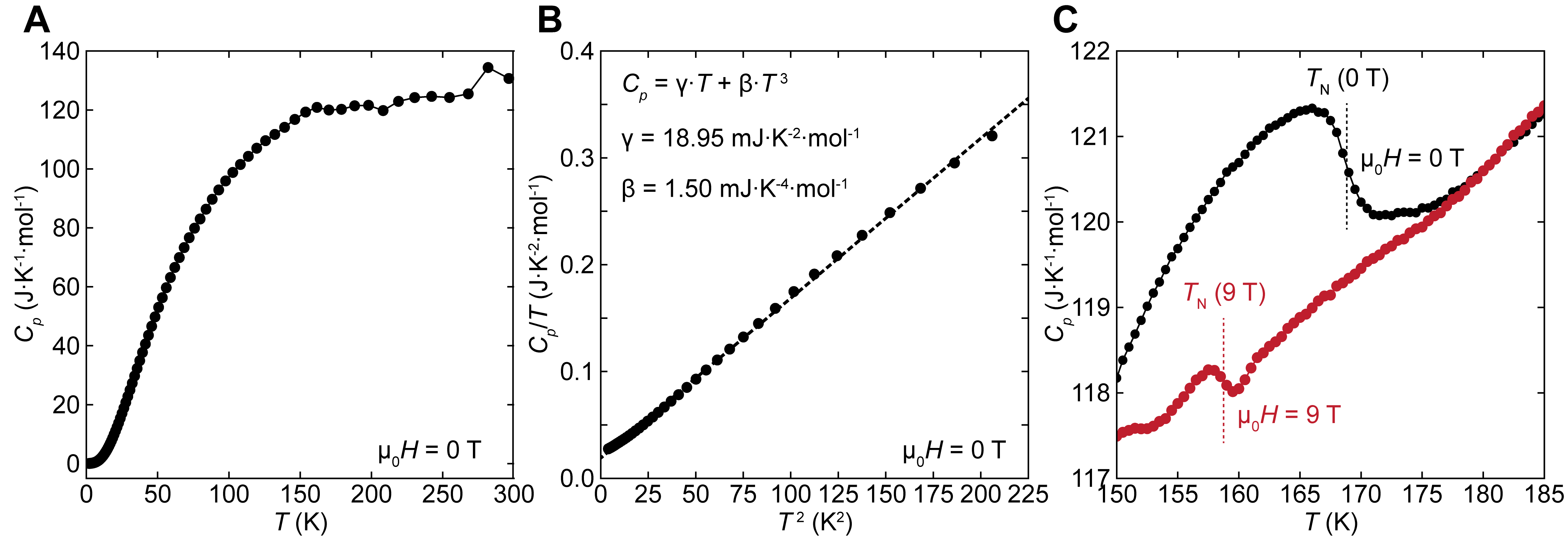}% Here is how to import EPS art
\centering
\caption[Specific heat capacity of TaFe\textsubscript{1.14}Te\textsubscript{3}]{\label{fig:figures5} Specific heat capacity of TaFe\textsubscript{1.14}Te\textsubscript{3}. A) Specific heat capacity versus temperature at zero magnetic field. B) Plot of specific heat capacity divided by the temperature versus the temperature squared. The dashed black line is a linear fit to the data. The fit equation and extracted fit parameters are given in the inset. C) Specific heat capacity versus temperature across the PM-AFM transition at zero magnetic field (solid black dots) and 9 T (solid red dots). The Néel temperature at 0 T and 9 T is denoted by a black and red dashed line, respectively.}
\end{figure*}
%%%%%%%%%%%%%%%%%%%%%%%%%%%%%%%%%%%%%%%%%%%%%%%%%%%%%%%%%%%%%%%%%%%%%%%%%%%%%%%%%%%%%%%%%%%%%%%%%%%%%%%%%%%%
%%%%%%%%%%%%%%%%%%%%%%%%%%%%%%%%%%%%%%%%%%%%%%%%%%%%%%%%%%%%%%%%%%%%%%%%%%%%%%%%%%%%%%%%%%%%%%%%%%%%%%%%%%%%

%%%%%%%%%%%%%%%%%%%%%%%%%%%%%%%%%%%%%%%%%%%%%%%%%%%%%%%%%%%%%%%%%%%%%%%%%%%%%%%%%%%%%%%%%%%%%%%%%%%%%%%%%%%%
%Figure s6
%%%%%%%%%%%%%%%%%%%%%%%%%%%%%%%%%%%%%%%%%%%%%%%%%%%%%%%%%%%%%%%%%%%%%%%%%%%%%%%%%%%%%%%%%%%%%%%%%%%%%%%%%%%%
\begin{figure*}[b]
\includegraphics[width=6.9in]{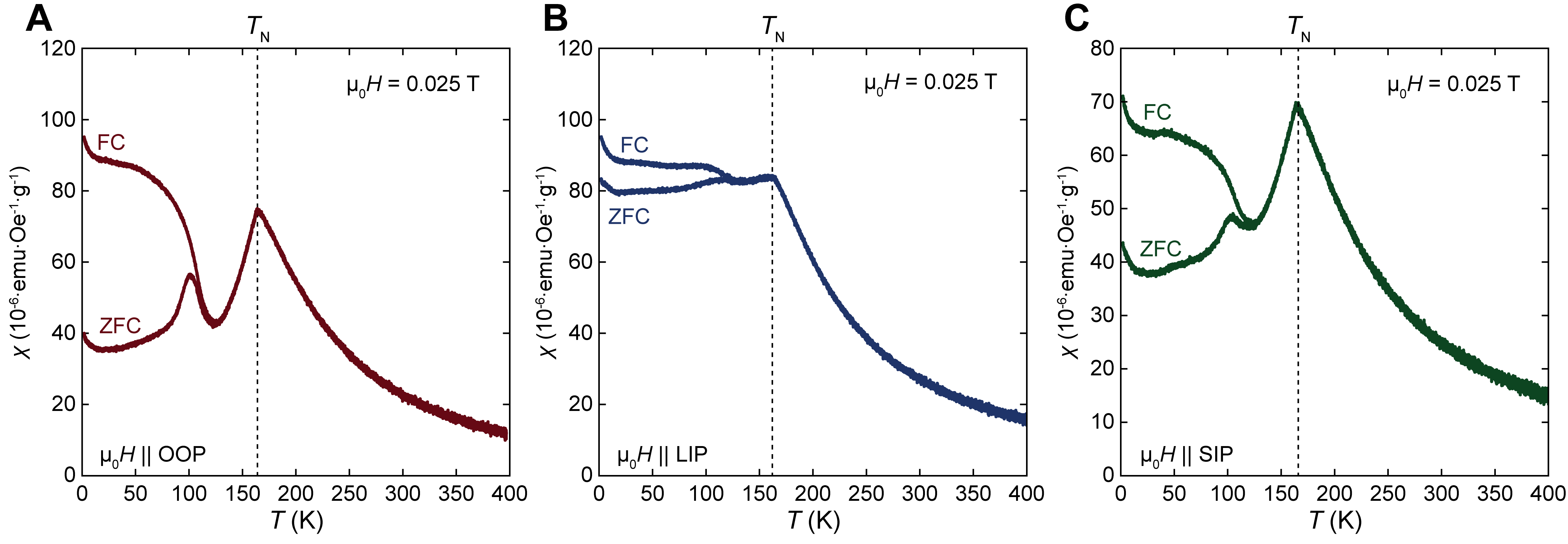}% Here is how to import EPS art
\centering
\caption[Observation of the spin-glass phase in TaFe\textsubscript{1.14}Te\textsubscript{3}]{\label{fig:figures6} Observation of the spin-glass phase in TaFe\textsubscript{1.14}Te\textsubscript{3}. A-C) Magnetic susceptibility versus temperature for magnetic fields oriented along the out-of-plane (A), long in-plane (B), and short-in-plane (C) directions. Both field-cooled and zero-field-cooled traces are shown. All traces were acquired with a magnetic field of 25 mT. \textit{T}\textsubscript{N} is denoted by a dashed black line in each plot.}
\end{figure*}
%%%%%%%%%%%%%%%%%%%%%%%%%%%%%%%%%%%%%%%%%%%%%%%%%%%%%%%%%%%%%%%%%%%%%%%%%%%%%%%%%%%%%%%%%%%%%%%%%%%%%%%%%%%%
%%%%%%%%%%%%%%%%%%%%%%%%%%%%%%%%%%%%%%%%%%%%%%%%%%%%%%%%%%%%%%%%%%%%%%%%%%%%%%%%%%%%%%%%%%%%%%%%%%%%%%%%%%%%

%%%%%%%%%%%%%%%%%%%%%%%%%%%%%%%%%%%%%%%%%%%%%%%%%%%%%%%%%%%%%%%%%%%%%%%%%%%%%%%%%%%%%%%%%%%%%%%%%%%%%%%%%%%%
%Figure s7
%%%%%%%%%%%%%%%%%%%%%%%%%%%%%%%%%%%%%%%%%%%%%%%%%%%%%%%%%%%%%%%%%%%%%%%%%%%%%%%%%%%%%%%%%%%%%%%%%%%%%%%%%%%%
\begin{figure*}[t]
\includegraphics[width=6.9in]{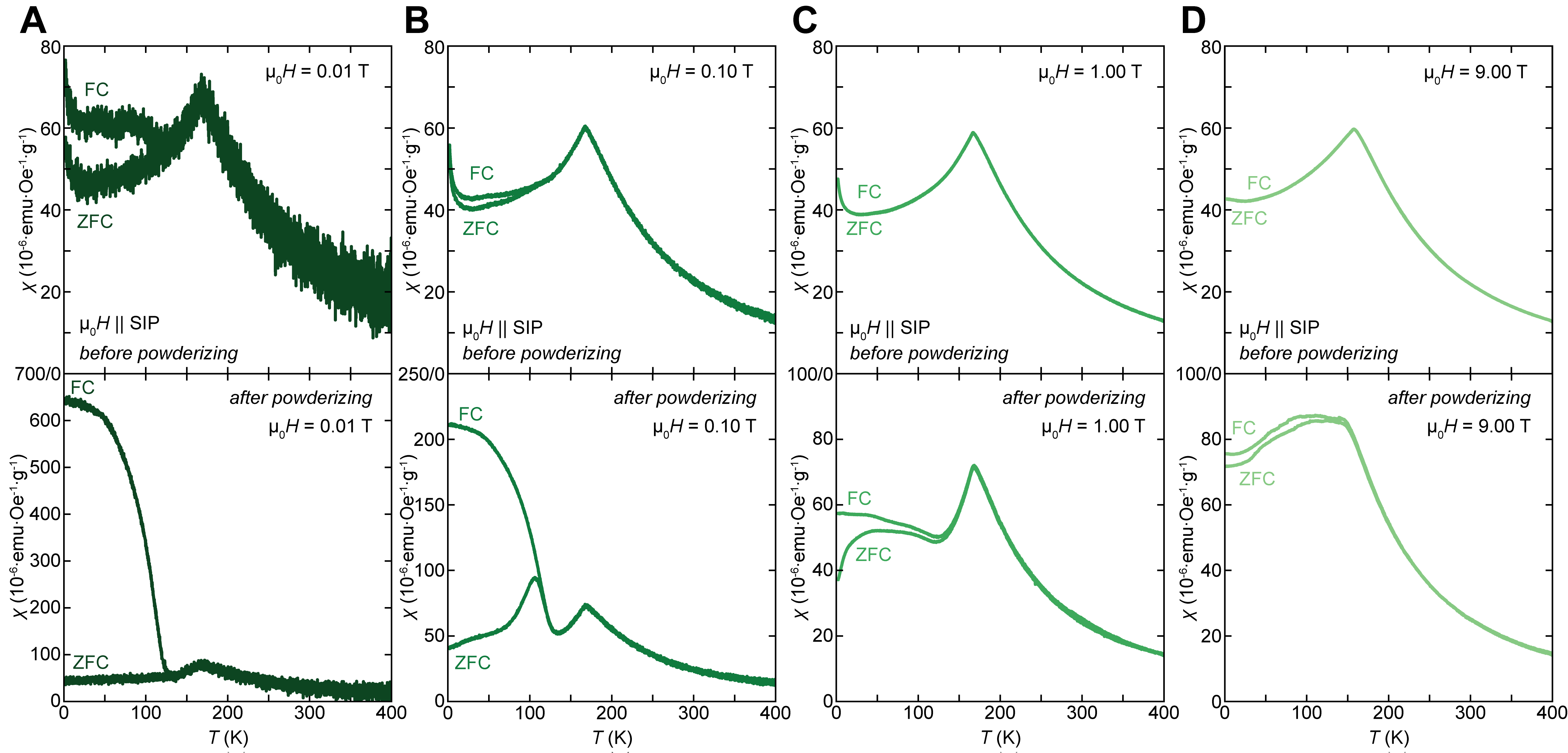}% Here is how to import EPS art
\centering
\caption[Powderization enhancement of the spin-glass phase in TaFe\textsubscript{1.14}Te\textsubscript{3}]{\label{fig:figures7} Powderization enhancement of the spin-glass phase in TaFe\textsubscript{1.14}Te\textsubscript{3}. A-D) Magnetic susceptibility versus temperature with a measuring field of 0.01 T (A), 0.1 T (B), 1 T (C), and 9 T (D). Both field-cooled and zero-field-cooled traces are shown and denoted. Top panels are collected from a single TaFe\textsubscript{1.14}Te\textsubscript{3} crystal with the magnetic field applied parallel to the short in-plane direction. Bottom panels are collected from the same powderized crystal.}
\end{figure*}
%%%%%%%%%%%%%%%%%%%%%%%%%%%%%%%%%%%%%%%%%%%%%%%%%%%%%%%%%%%%%%%%%%%%%%%%%%%%%%%%%%%%%%%%%%%%%%%%%%%%%%%%%%%%
%%%%%%%%%%%%%%%%%%%%%%%%%%%%%%%%%%%%%%%%%%%%%%%%%%%%%%%%%%%%%%%%%%%%%%%%%%%%%%%%%%%%%%%%%%%%%%%%%%%%%%%%%%%%

%%%%%%%%%%%%%%%%%%%%%%%%%%%%%%%%%%%%%%%%%%%%%%%%%%%%%%%%%%%%%%%%%%%%%%%%%%%%%%%%%%%%%%%%%%%%%%%%%%%%%%%%%%%%
%Figure s8
%%%%%%%%%%%%%%%%%%%%%%%%%%%%%%%%%%%%%%%%%%%%%%%%%%%%%%%%%%%%%%%%%%%%%%%%%%%%%%%%%%%%%%%%%%%%%%%%%%%%%%%%%%%%
\begin{figure*}[b]
\includegraphics[width=6.9in]{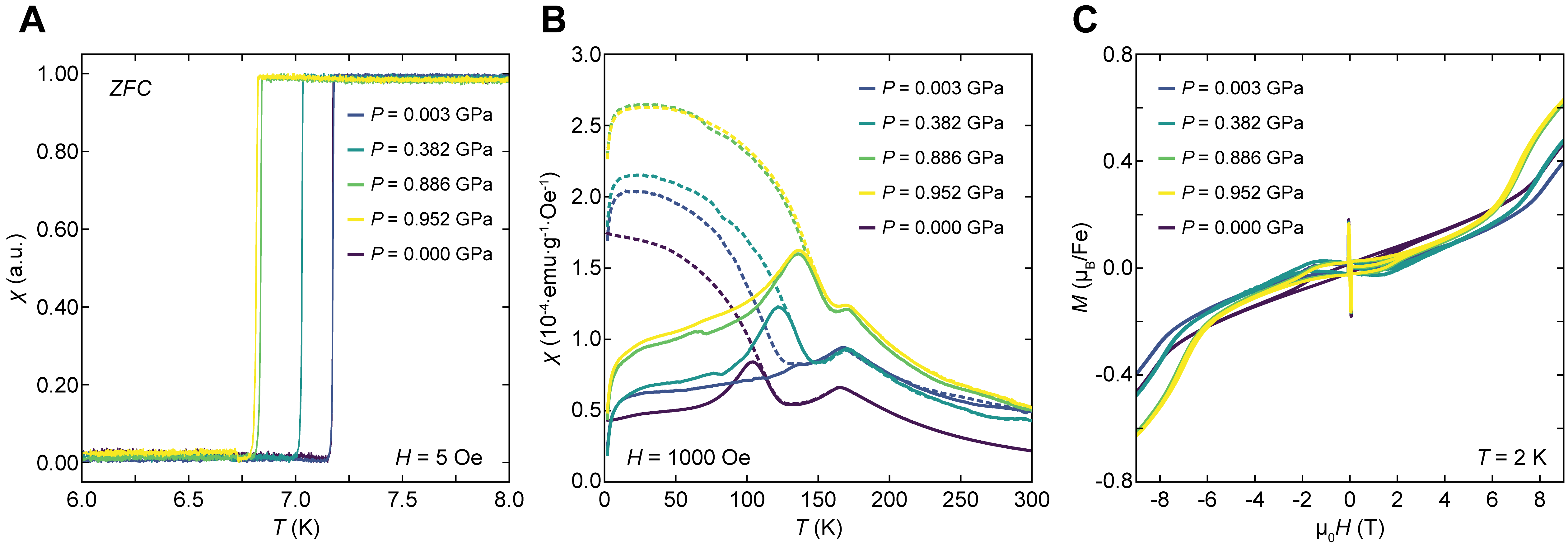}% Here is how to import EPS art
\centering
\caption[Magnetic properties of TaFe\textsubscript{1.14}Te\textsubscript{3} spin-glass phase under hydrostatic pressure]{\label{fig:figures8} Magnetic properties of TaFe\textsubscript{1.14}Te\textsubscript{3} spin-glass phase under hydrostatic pressure. A) Magnetic susceptibility versus temperature across the superconducting transition for the lead manometer for various applied pressures. The corresponding applied pressure is given in the inset. All traces were offset and normalized for clarity. A measuring field of 5 Oe was used. B) Field-cooled (dashed traces) and zero-field-cooled (solid traces) magnetic susceptibility versus temperature for various applied pressures. The corresponding pressure is given in the inset. A measuring field of 1000 Oe was used. C) Magnetization versus magnetic field at 2 K for various applied pressures. Due to the pressure-cell preparation, the magnetic field direction is averaged along all crystal axes. The corresponding pressure is given in the inset. For all panels, measurements were performed in order of increasing pressure. The final measurement (0.000 GPa) is a reference measurement of the sample plus Teflon capsule measured outside of the pressure cell on a brass paddle sample mount. }
\end{figure*}
%%%%%%%%%%%%%%%%%%%%%%%%%%%%%%%%%%%%%%%%%%%%%%%%%%%%%%%%%%%%%%%%%%%%%%%%%%%%%%%%%%%%%%%%%%%%%%%%%%%%%%%%%%%%
%%%%%%%%%%%%%%%%%%%%%%%%%%%%%%%%%%%%%%%%%%%%%%%%%%%%%%%%%%%%%%%%%%%%%%%%%%%%%%%%%%%%%%%%%%%%%%%%%%%%%%%%%%%%

%%%%%%%%%%%%%%%%%%%%%%%%%%%%%%%%%%%%%%%%%%%%%%%%%%%%%%%%%%%%%%%%%%%%%%%%%%%%%%%%%%%%%%%%%%%%%%%%%%%%%%%%%%%%
%Figure s9
%%%%%%%%%%%%%%%%%%%%%%%%%%%%%%%%%%%%%%%%%%%%%%%%%%%%%%%%%%%%%%%%%%%%%%%%%%%%%%%%%%%%%%%%%%%%%%%%%%%%%%%%%%%%
\begin{figure*}[t]
\includegraphics[width=6.9in]{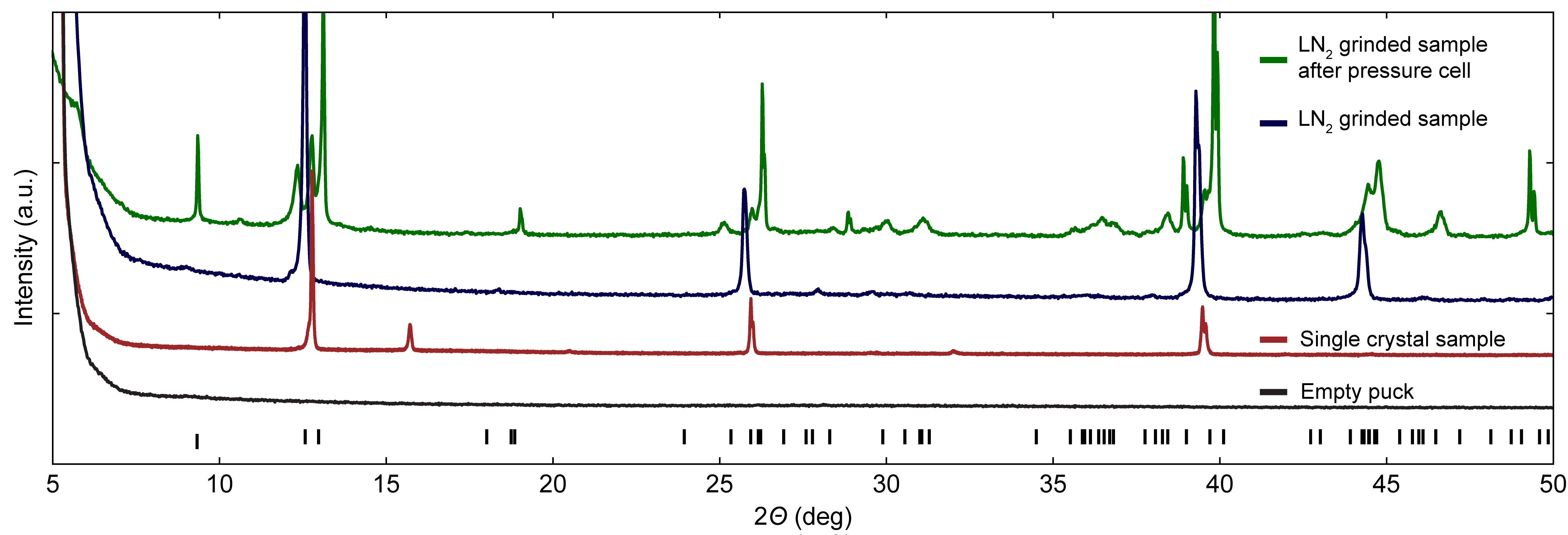}% Here is how to import EPS art
\centering
\caption[Powder X-ray diffraction of bulk TaFe\textsubscript{1.14}Te\textsubscript{3} crystals before and after grinding]{\label{fig:figures9} Powder X-ray diffraction of bulk TaFe\textsubscript{1.14}Te\textsubscript{3} crystals before and after grinding. Powder X-ray diffraction (PXRD) intensity versus 2\texttheta{ }for the empty puck (solid black line), a single crystal of TaFe\textsubscript{1.14}Te\textsubscript{3} (solid red line), a TaFe\textsubscript{1.14}Te\textsubscript{3} single crystal ground with LN\textsubscript{2} (solid blue line), and a TaFe\textsubscript{1.14}Te\textsubscript{3} single crystal ground with LN\textsubscript{2} after being pressurized up to 1.3 GPa (solid green line). Simulated TaFe\textsubscript{1.14}Te\textsubscript{3} PXRD peaks are denoted with black tick marks.}
\end{figure*}
%%%%%%%%%%%%%%%%%%%%%%%%%%%%%%%%%%%%%%%%%%%%%%%%%%%%%%%%%%%%%%%%%%%%%%%%%%%%%%%%%%%%%%%%%%%%%%%%%%%%%%%%%%%%
%%%%%%%%%%%%%%%%%%%%%%%%%%%%%%%%%%%%%%%%%%%%%%%%%%%%%%%%%%%%%%%%%%%%%%%%%%%%%%%%%%%%%%%%%%%%%%%%%%%%%%%%%%%%

%%%%%%%%%%%%%%%%%%%%%%%%%%%%%%%%%%%%%%%%%%%%%%%%%%%%%%%%%%%%%%%%%%%%%%%%%%%%%%%%%%%%%%%%%%%%%%%%%%%%%%%%%%%%
%Figure s10
%%%%%%%%%%%%%%%%%%%%%%%%%%%%%%%%%%%%%%%%%%%%%%%%%%%%%%%%%%%%%%%%%%%%%%%%%%%%%%%%%%%%%%%%%%%%%%%%%%%%%%%%%%%%
\begin{figure*}[b]
\includegraphics[width=6.9in]{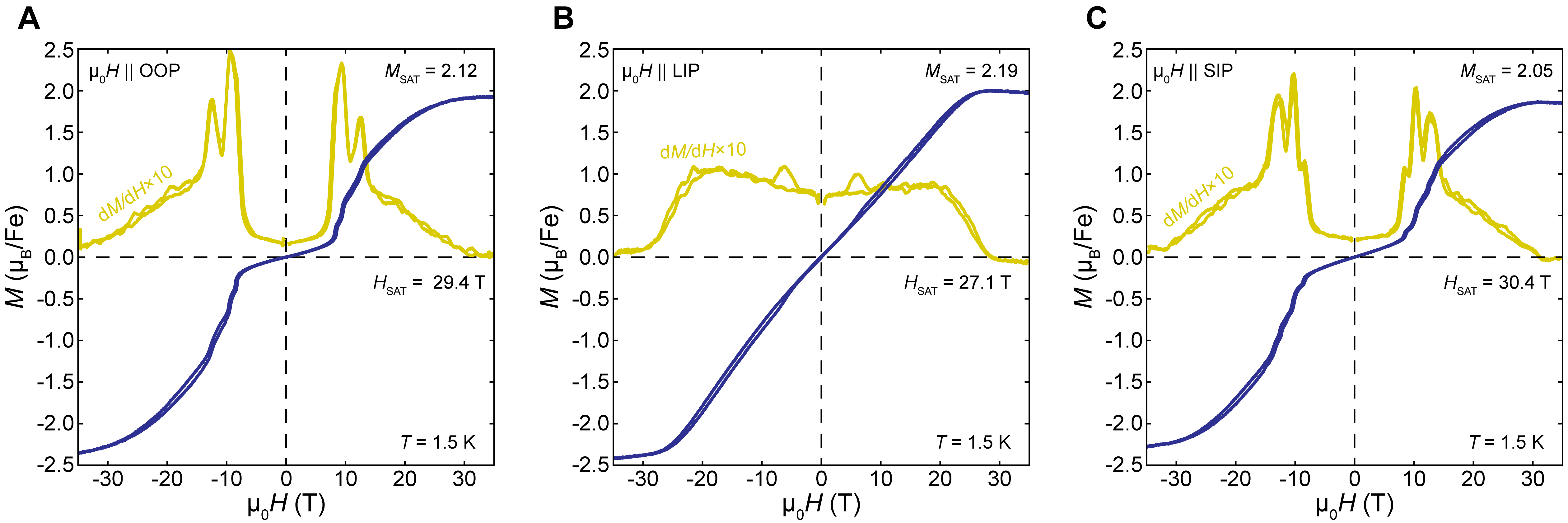}% Here is how to import EPS art
\centering
\caption[Analysis of TaFe\textsubscript{1.14}Te\textsubscript{3} high-field magnetometry at 1.5 K]{\label{fig:figures10} Analysis of TaFe\textsubscript{1.14}Te\textsubscript{3} high-field magnetometry at 1.5 K. A-C) Magnetization (solid blue lines) and derivative of magnetization (solid yellow lines) versus magnetic field at 1.5 K for magnetic fields oriented along the out-of-plane (A), long in-plane (B), and short in-plane (C) directions. In each plot, the extracted saturation magnetization and saturation magnetic field are given in the inset.}
\end{figure*}
%%%%%%%%%%%%%%%%%%%%%%%%%%%%%%%%%%%%%%%%%%%%%%%%%%%%%%%%%%%%%%%%%%%%%%%%%%%%%%%%%%%%%%%%%%%%%%%%%%%%%%%%%%%%
%%%%%%%%%%%%%%%%%%%%%%%%%%%%%%%%%%%%%%%%%%%%%%%%%%%%%%%%%%%%%%%%%%%%%%%%%%%%%%%%%%%%%%%%%%%%%%%%%%%%%%%%%%%%

%%%%%%%%%%%%%%%%%%%%%%%%%%%%%%%%%%%%%%%%%%%%%%%%%%%%%%%%%%%%%%%%%%%%%%%%%%%%%%%%%%%%%%%%%%%%%%%%%%%%%%%%%%%%
%Figure s11
%%%%%%%%%%%%%%%%%%%%%%%%%%%%%%%%%%%%%%%%%%%%%%%%%%%%%%%%%%%%%%%%%%%%%%%%%%%%%%%%%%%%%%%%%%%%%%%%%%%%%%%%%%%%
\begin{figure*}[t]
\includegraphics[width=6.9in]{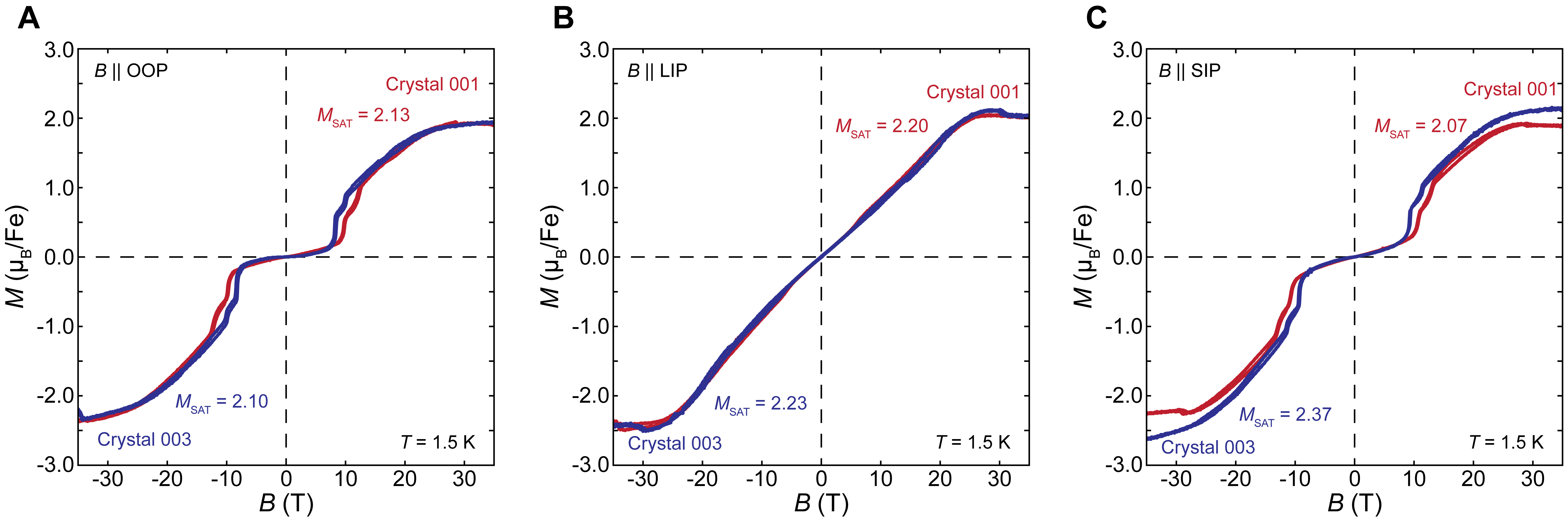}% Here is how to import EPS art
\centering
\caption[High-field magnetometry on two TaFe\textsubscript{1.14}Te\textsubscript{3} single crystals]{\label{fig:figures11} High-field magnetometry on two TaFe\textsubscript{1.14}Te\textsubscript{3} single crystals. A-C) Magnetization versus magnetic field at 1.5 K for magnetic fields oriented along the out-of-plane (A), long in-plane (B), and short in-plane (C) directions for two separate TaFe\textsubscript{1.14}Te\textsubscript{3} crystals. Each plot's blue and red traces correspond to the different TaFe\textsubscript{1.14}Te\textsubscript{3} crystals. In each plot, the extracted saturation magnetization is given in the inset.}
\end{figure*}
%%%%%%%%%%%%%%%%%%%%%%%%%%%%%%%%%%%%%%%%%%%%%%%%%%%%%%%%%%%%%%%%%%%%%%%%%%%%%%%%%%%%%%%%%%%%%%%%%%%%%%%%%%%%
%%%%%%%%%%%%%%%%%%%%%%%%%%%%%%%%%%%%%%%%%%%%%%%%%%%%%%%%%%%%%%%%%%%%%%%%%%%%%%%%%%%%%%%%%%%%%%%%%%%%%%%%%%%%

%%%%%%%%%%%%%%%%%%%%%%%%%%%%%%%%%%%%%%%%%%%%%%%%%%%%%%%%%%%%%%%%%%%%%%%%%%%%%%%%%%%%%%%%%%%%%%%%%%%%%%%%%%%%
%Figure s12
%%%%%%%%%%%%%%%%%%%%%%%%%%%%%%%%%%%%%%%%%%%%%%%%%%%%%%%%%%%%%%%%%%%%%%%%%%%%%%%%%%%%%%%%%%%%%%%%%%%%%%%%%%%%
\begin{figure*}[b]
\includegraphics[width=6.9in]{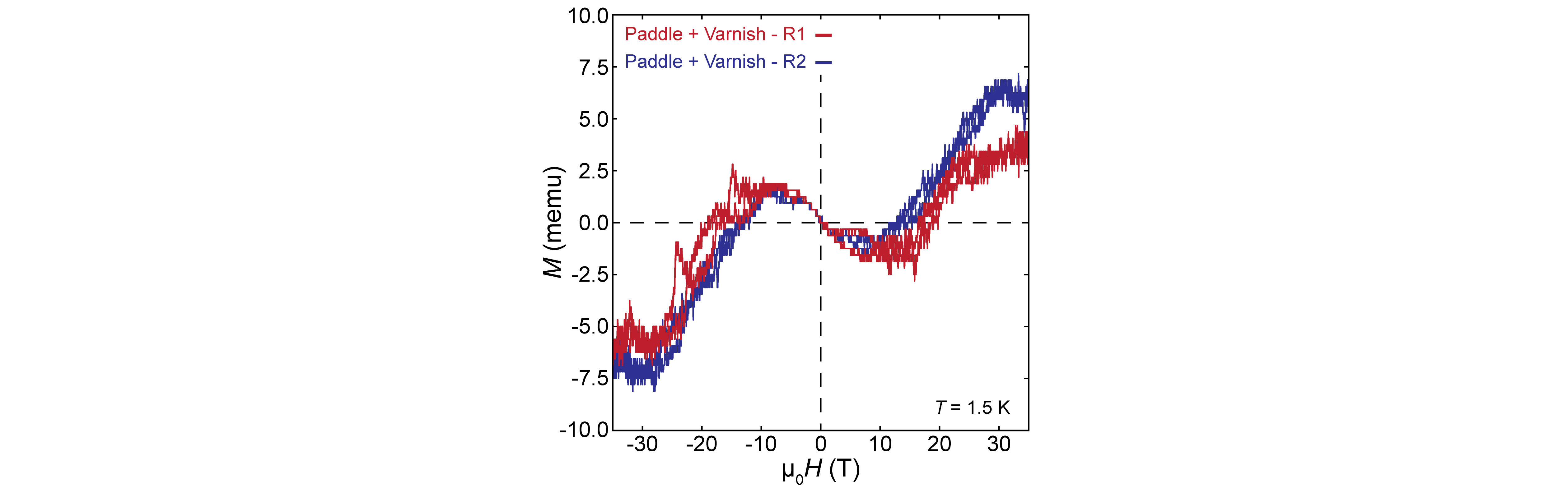}% Here is how to import EPS art
\centering
\caption[Background high-field magnetization from the sample paddle]{\label{fig:figures12} Background high-field magnetization from the sample paddle. Magnetization versus magnetic field at 1.5 K for two sample paddles with GE varnish. The blue and red traces correspond to the different paddles.}
\end{figure*}
%%%%%%%%%%%%%%%%%%%%%%%%%%%%%%%%%%%%%%%%%%%%%%%%%%%%%%%%%%%%%%%%%%%%%%%%%%%%%%%%%%%%%%%%%%%%%%%%%%%%%%%%%%%%
%%%%%%%%%%%%%%%%%%%%%%%%%%%%%%%%%%%%%%%%%%%%%%%%%%%%%%%%%%%%%%%%%%%%%%%%%%%%%%%%%%%%%%%%%%%%%%%%%%%%%%%%%%%%

%%%%%%%%%%%%%%%%%%%%%%%%%%%%%%%%%%%%%%%%%%%%%%%%%%%%%%%%%%%%%%%%%%%%%%%%%%%%%%%%%%%%%%%%%%%%%%%%%%%%%%%%%%%%
%Figure s13
%%%%%%%%%%%%%%%%%%%%%%%%%%%%%%%%%%%%%%%%%%%%%%%%%%%%%%%%%%%%%%%%%%%%%%%%%%%%%%%%%%%%%%%%%%%%%%%%%%%%%%%%%%%%
\begin{figure*}[t]
\includegraphics[width=6.9in]{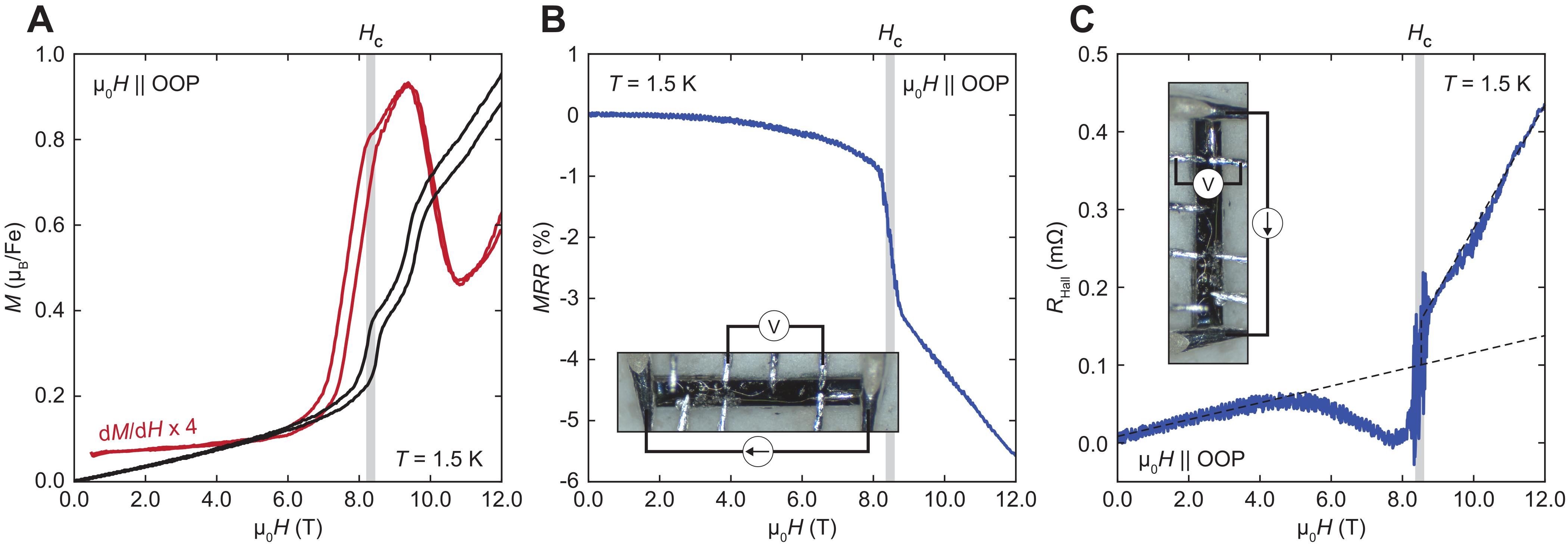}% Here is how to import EPS art
\centering
\caption[Analysis of the metamagnetic transition far below \textit{T}\textsubscript{N} in TaFe\textsubscript{1.14}Te\textsubscript{3}]{\label{fig:figures13} Analysis of the metamagnetic transition far below \textit{T}\textsubscript{N} in TaFe\textsubscript{1.14}Te\textsubscript{3}. A) Magnetization (solid black line) and the derivative of magnetization (solid red line) versus magnetic field at 1.5 K for fields parallel to the out-of-plane direction. B) Magnetoresistance ratio, \textit{MRR} = [\textit{R}(\textit{H})-\textit{R}(\textit{H}=0)]/\textit{R}(\textit{H}=0)$\times$100, versus magnetic field at 1.5 K for fields parallel to the out-of-plane direction. The inset is an optical image of the device along with the measurement configuration. C) Hall resistance versus magnetic field at 1.5 K. Two distinct linear regions (denoted by dashed black lines) are observed below \textit{T}\textsubscript{N}. The inset is an optical image of the device along with the measurement configuration. In A-C, the vertical gray bars demarcate the magnetic field at which a spin-flop metamagnetic transition occurs (\textit{H}\textsubscript{C}). This spin-flop transition (A) results in a sharp decrease in \textit{MRR} (B) and a sharp increase in the anomalous Hall contribution to \textit{R}\textsubscript{Hall} (C).}
\end{figure*}
%%%%%%%%%%%%%%%%%%%%%%%%%%%%%%%%%%%%%%%%%%%%%%%%%%%%%%%%%%%%%%%%%%%%%%%%%%%%%%%%%%%%%%%%%%%%%%%%%%%%%%%%%%%%
%%%%%%%%%%%%%%%%%%%%%%%%%%%%%%%%%%%%%%%%%%%%%%%%%%%%%%%%%%%%%%%%%%%%%%%%%%%%%%%%%%%%%%%%%%%%%%%%%%%%%%%%%%%%

%%%%%%%%%%%%%%%%%%%%%%%%%%%%%%%%%%%%%%%%%%%%%%%%%%%%%%%%%%%%%%%%%%%%%%%%%%%%%%%%%%%%%%%%%%%%%%%%%%%%%%%%%%%%
%Figure s14
%%%%%%%%%%%%%%%%%%%%%%%%%%%%%%%%%%%%%%%%%%%%%%%%%%%%%%%%%%%%%%%%%%%%%%%%%%%%%%%%%%%%%%%%%%%%%%%%%%%%%%%%%%%%
\begin{figure*}[b]
\includegraphics[width=6.9in]{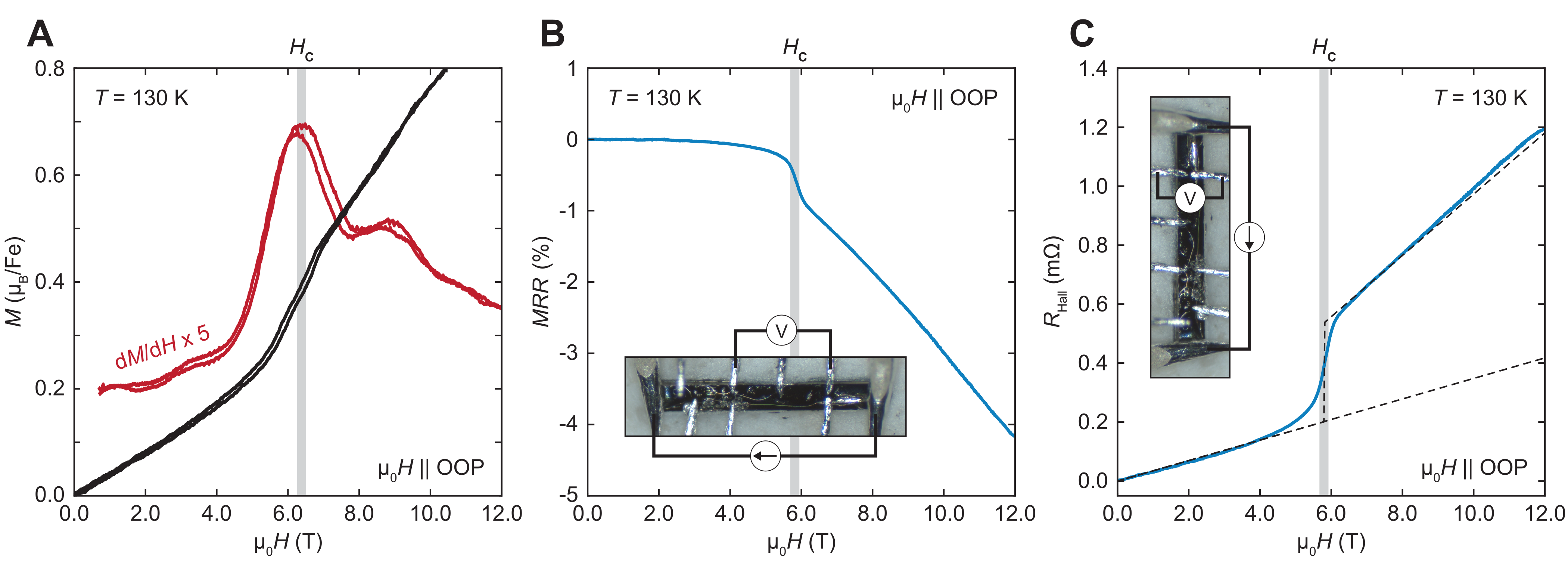}% Here is how to import EPS art
\centering
\caption[Analysis of the metamagnetic transition near \textit{T}\textsubscript{N} in TaFe\textsubscript{1.14}Te\textsubscript{3}]{\label{fig:figures14} Analysis of the metamagnetic transition near \textit{T}\textsubscript{N} in TaFe\textsubscript{1.14}Te\textsubscript{3}. A) Magnetization (solid black line) and the derivative of magnetization (solid red line) versus magnetic field at 130 K for fields parallel to the out-of-plane direction. B) Magnetoresistance ratio, \textit{MRR} = [\textit{R}(\textit{H})-\textit{R}(\textit{H}=0)]/\textit{R}(\textit{H}=0)$\times$100, versus magnetic field at 130 K for fields parallel to the out-of-plane direction. The inset is an optical image of the device along with the measurement configuration. C) Hall resistance versus magnetic field at 130 K. Two distinct linear regions (denoted by dashed black lines) are observed below \textit{T}\textsubscript{N}. The inset is an optical image of the device along with the measurement configuration. In A-C, the vertical gray bars demarcate the magnetic field at which a spin-flop metamagnetic transition occurs (\textit{H}\textsubscript{C}). This spin-flop transition (A) results in a sharp decrease in MRR (B) and a sharp increase in the anomalous Hall contribution to \textit{R}\textsubscript{Hall} (C).}
\end{figure*}
%%%%%%%%%%%%%%%%%%%%%%%%%%%%%%%%%%%%%%%%%%%%%%%%%%%%%%%%%%%%%%%%%%%%%%%%%%%%%%%%%%%%%%%%%%%%%%%%%%%%%%%%%%%%
%%%%%%%%%%%%%%%%%%%%%%%%%%%%%%%%%%%%%%%%%%%%%%%%%%%%%%%%%%%%%%%%%%%%%%%%%%%%%%%%%%%%%%%%%%%%%%%%%%%%%%%%%%%%

%%%%%%%%%%%%%%%%%%%%%%%%%%%%%%%%%%%%%%%%%%%%%%%%%%%%%%%%%%%%%%%%%%%%%%%%%%%%%%%%%%%%%%%%%%%%%%%%%%%%%%%%%%%%
%Figure s15
%%%%%%%%%%%%%%%%%%%%%%%%%%%%%%%%%%%%%%%%%%%%%%%%%%%%%%%%%%%%%%%%%%%%%%%%%%%%%%%%%%%%%%%%%%%%%%%%%%%%%%%%%%%%
\begin{figure*}[t]
\includegraphics[width=6.9in]{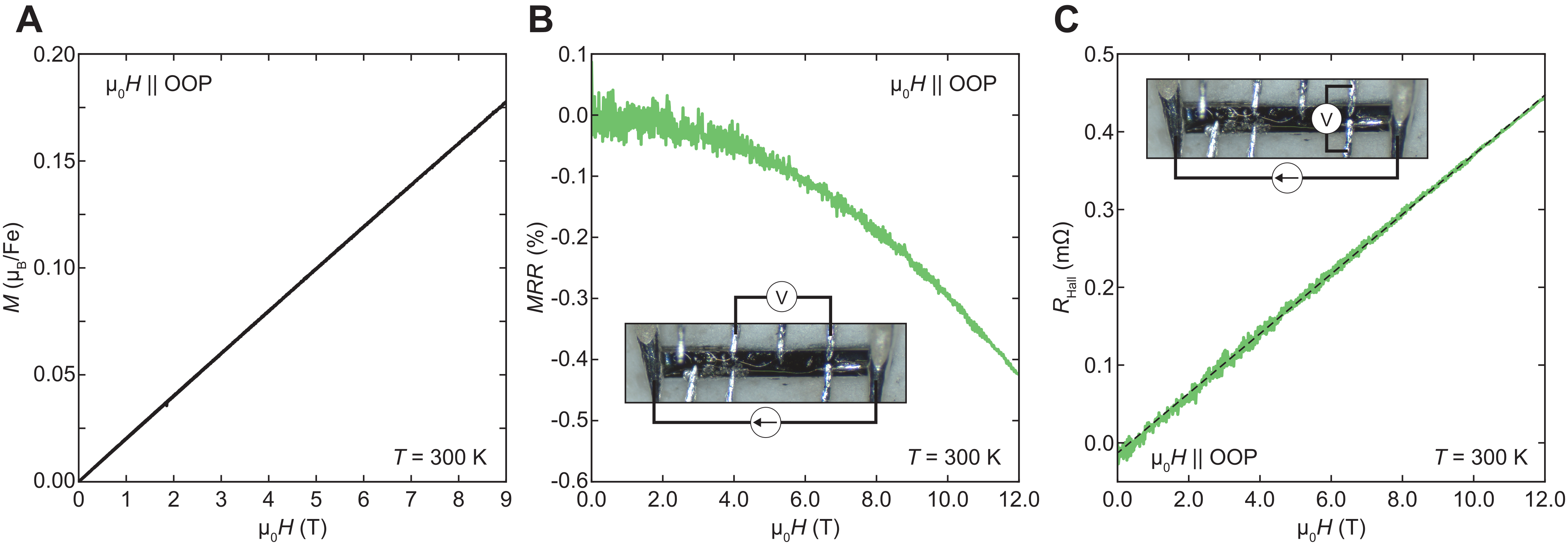}% Here is how to import EPS art
\centering
\caption[Analysis of magnetometry and transport above \textit{T}\textsubscript{N} in TaFe\textsubscript{1.14}Te\textsubscript{3}]{\label{fig:figures15} Analysis of magnetometry and transport above \textit{T}\textsubscript{N} in TaFe\textsubscript{1.14}Te\textsubscript{3}. A) Magnetization (solid black line) versus magnetic field at 300 K for fields parallel to the out-of-plane direction. B) Magnetoresistance ratio, \textit{MRR} = [\textit{R}(\textit{H})-\textit{R}(\textit{H}=0)]/\textit{R}(\textit{H}=0)$\times$100, versus magnetic field at 300 K for fields parallel to the out-of-plane direction. The inset is an optical image of the device along with the measurement configuration. C) Hall resistance versus magnetic field at 300 K. A single linear region (denoted by a dashed black line) is observed over the entire field range. The inset is an optical image of the device along with the measurement configuration.}
\end{figure*}
%%%%%%%%%%%%%%%%%%%%%%%%%%%%%%%%%%%%%%%%%%%%%%%%%%%%%%%%%%%%%%%%%%%%%%%%%%%%%%%%%%%%%%%%%%%%%%%%%%%%%%%%%%%%
%%%%%%%%%%%%%%%%%%%%%%%%%%%%%%%%%%%%%%%%%%%%%%%%%%%%%%%%%%%%%%%%%%%%%%%%%%%%%%%%%%%%%%%%%%%%%%%%%%%%%%%%%%%%

%%%%%%%%%%%%%%%%%%%%%%%%%%%%%%%%%%%%%%%%%%%%%%%%%%%%%%%%%%%%%%%%%%%%%%%%%%%%%%%%%%%%%%%%%%%%%%%%%%%%%%%%%%%%
%Figure s16
%%%%%%%%%%%%%%%%%%%%%%%%%%%%%%%%%%%%%%%%%%%%%%%%%%%%%%%%%%%%%%%%%%%%%%%%%%%%%%%%%%%%%%%%%%%%%%%%%%%%%%%%%%%%
\begin{figure*}[b]
\includegraphics[width=6.9in]{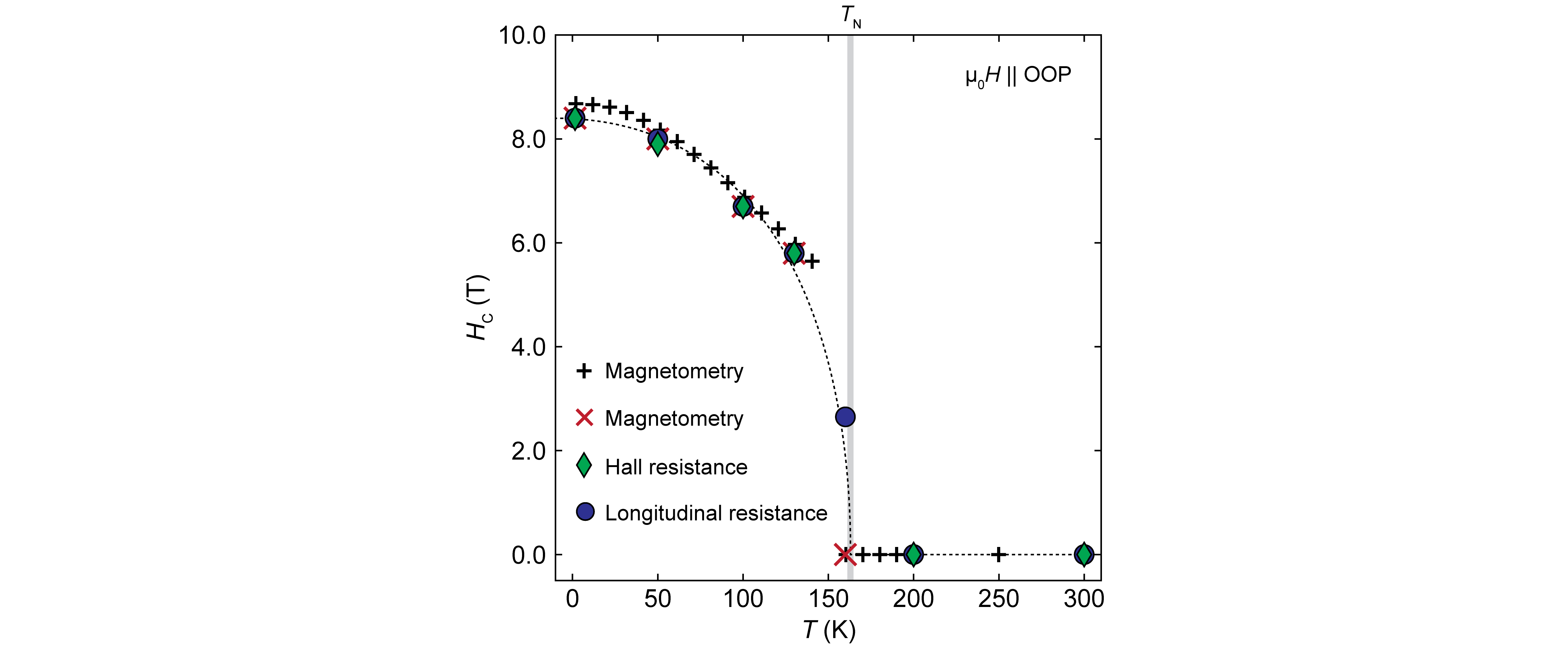}% Here is how to import EPS art
\centering
\caption[Temperature dependence of the first metamagnetic transition field in TaFe\textsubscript{1.14}Te\textsubscript{3}]{\label{fig:figures16} Temperature dependence of the first metamagnetic transition field in TaFe\textsubscript{1.14}Te\textsubscript{3}. The magnetic field at which the first metamagnetic transition occurs as measured by magnetometry (solid black pluses and red crosses), longitudinal magnetoresistance (solid blue circles), and Hall resistance (solid green diamonds). The dashed black line is a guide to the eye. The Néel temperature is denoted by a solid gray line. The magnetic field was aligned parallel to the out-of-plane axis for all measurements.}
\end{figure*}
%%%%%%%%%%%%%%%%%%%%%%%%%%%%%%%%%%%%%%%%%%%%%%%%%%%%%%%%%%%%%%%%%%%%%%%%%%%%%%%%%%%%%%%%%%%%%%%%%%%%%%%%%%%%
%%%%%%%%%%%%%%%%%%%%%%%%%%%%%%%%%%%%%%%%%%%%%%%%%%%%%%%%%%%%%%%%%%%%%%%%%%%%%%%%%%%%%%%%%%%%%%%%%%%%%%%%%%%%

%%%%%%%%%%%%%%%%%%%%%%%%%%%%%%%%%%%%%%%%%%%%%%%%%%%%%%%%%%%%%%%%%%%%%%%%%%%%%%%%%%%%%%%%%%%%%%%%%%%%%%%%%%%%
%Figure s17
%%%%%%%%%%%%%%%%%%%%%%%%%%%%%%%%%%%%%%%%%%%%%%%%%%%%%%%%%%%%%%%%%%%%%%%%%%%%%%%%%%%%%%%%%%%%%%%%%%%%%%%%%%%%
\begin{figure*}[t]
\includegraphics[width=6.9in]{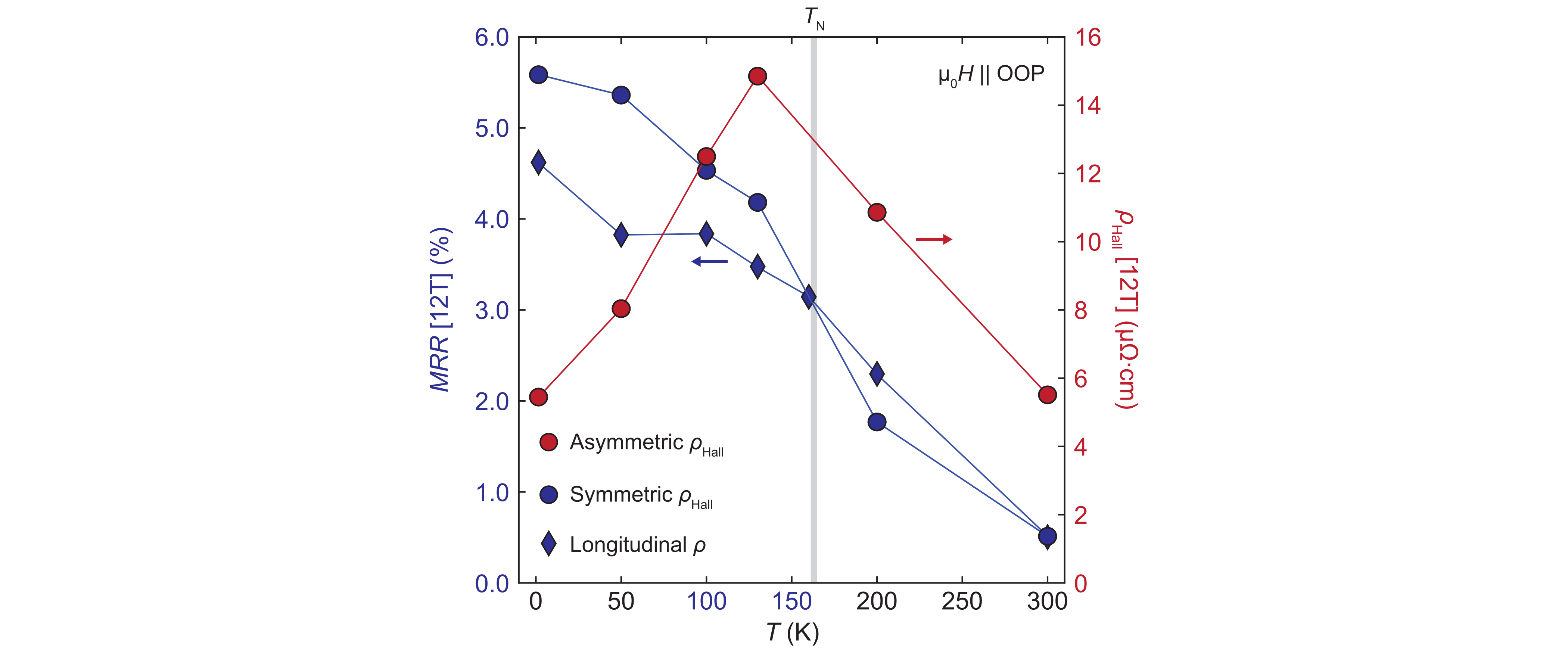}% Here is how to import EPS art
\centering
\caption[Temperature dependence of the 12 T \textit{MRR} and Hall resistance in TaFe\textsubscript{1.14}Te\textsubscript{3}]{\label{fig:figures17} Temperature dependence of the 12 T \textit{MRR} and Hall resistance in TaFe\textsubscript{1.14}Te\textsubscript{3}. Left axis: \textit{MRR} at 12 T versus temperature extracted from longitudinal resistance measurements (solid blue diamonds) and the symmetric component of the Hall resistance measurements (solid blue circles). Right axis: Hall resistance at 12 T versus temperature. The Hall resistance was anti-symmetrized to remove components of the raw data that were symmetric in the magnetic field. The Néel temperature is denoted by a solid gray line. The magnetic field was oriented parallel to the out-of-plane axis for all measurements.}
\end{figure*}
%%%%%%%%%%%%%%%%%%%%%%%%%%%%%%%%%%%%%%%%%%%%%%%%%%%%%%%%%%%%%%%%%%%%%%%%%%%%%%%%%%%%%%%%%%%%%%%%%%%%%%%%%%%%
%%%%%%%%%%%%%%%%%%%%%%%%%%%%%%%%%%%%%%%%%%%%%%%%%%%%%%%%%%%%%%%%%%%%%%%%%%%%%%%%%%%%%%%%%%%%%%%%%%%%%%%%%%%%

%%%%%%%%%%%%%%%%%%%%%%%%%%%%%%%%%%%%%%%%%%%%%%%%%%%%%%%%%%%%%%%%%%%%%%%%%%%%%%%%%%%%%%%%%%%%%%%%%%%%%%%%%%%%
%Figure s18
%%%%%%%%%%%%%%%%%%%%%%%%%%%%%%%%%%%%%%%%%%%%%%%%%%%%%%%%%%%%%%%%%%%%%%%%%%%%%%%%%%%%%%%%%%%%%%%%%%%%%%%%%%%%
\begin{figure*}[b]
\includegraphics[width=6.9in]{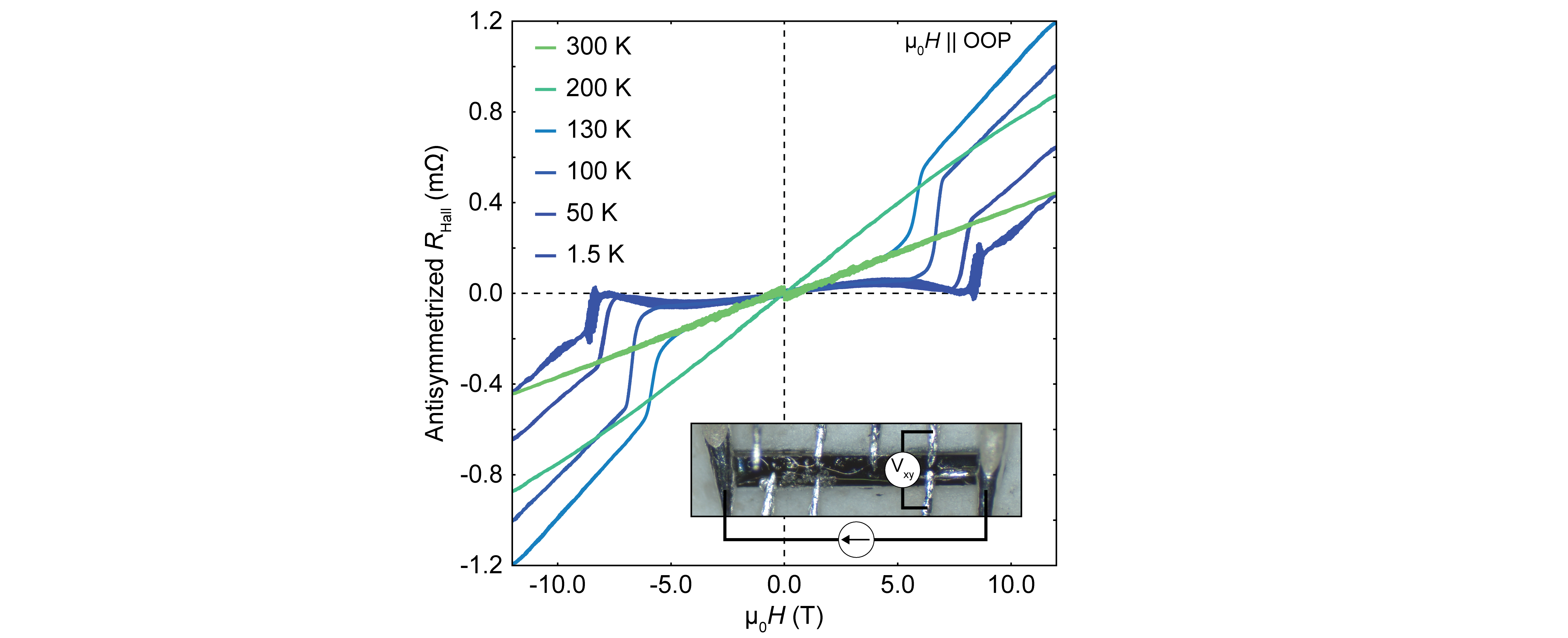}% Here is how to import EPS art
\centering
\caption[Temperature dependence of anti-symmetrized Hall data in TaFe\textsubscript{1.14}Te\textsubscript{3}]{\label{fig:figures18} Temperature dependence of anti-symmetrized Hall data in TaFe\textsubscript{1.14}Te\textsubscript{3}. Anti-symmetrized Hall resistance versus magnetic field at various temperatures for fields parallel to the out-of-plane direction. Both forward and backward traces are shown. The temperature at which each trace was taken is given in the inset. The dashed black lines denote the zero crossings. The inset is an optical image of the device along with the measurement configuration.}
\end{figure*}
%%%%%%%%%%%%%%%%%%%%%%%%%%%%%%%%%%%%%%%%%%%%%%%%%%%%%%%%%%%%%%%%%%%%%%%%%%%%%%%%%%%%%%%%%%%%%%%%%%%%%%%%%%%%
%%%%%%%%%%%%%%%%%%%%%%%%%%%%%%%%%%%%%%%%%%%%%%%%%%%%%%%%%%%%%%%%%%%%%%%%%%%%%%%%%%%%%%%%%%%%%%%%%%%%%%%%%%%%

%%%%%%%%%%%%%%%%%%%%%%%%%%%%%%%%%%%%%%%%%%%%%%%%%%%%%%%%%%%%%%%%%%%%%%%%%%%%%%%%%%%%%%%%%%%%%%%%%%%%%%%%%%%%
%Figure s19
%%%%%%%%%%%%%%%%%%%%%%%%%%%%%%%%%%%%%%%%%%%%%%%%%%%%%%%%%%%%%%%%%%%%%%%%%%%%%%%%%%%%%%%%%%%%%%%%%%%%%%%%%%%%
\begin{figure*}[t]
\includegraphics[width=6.9in]{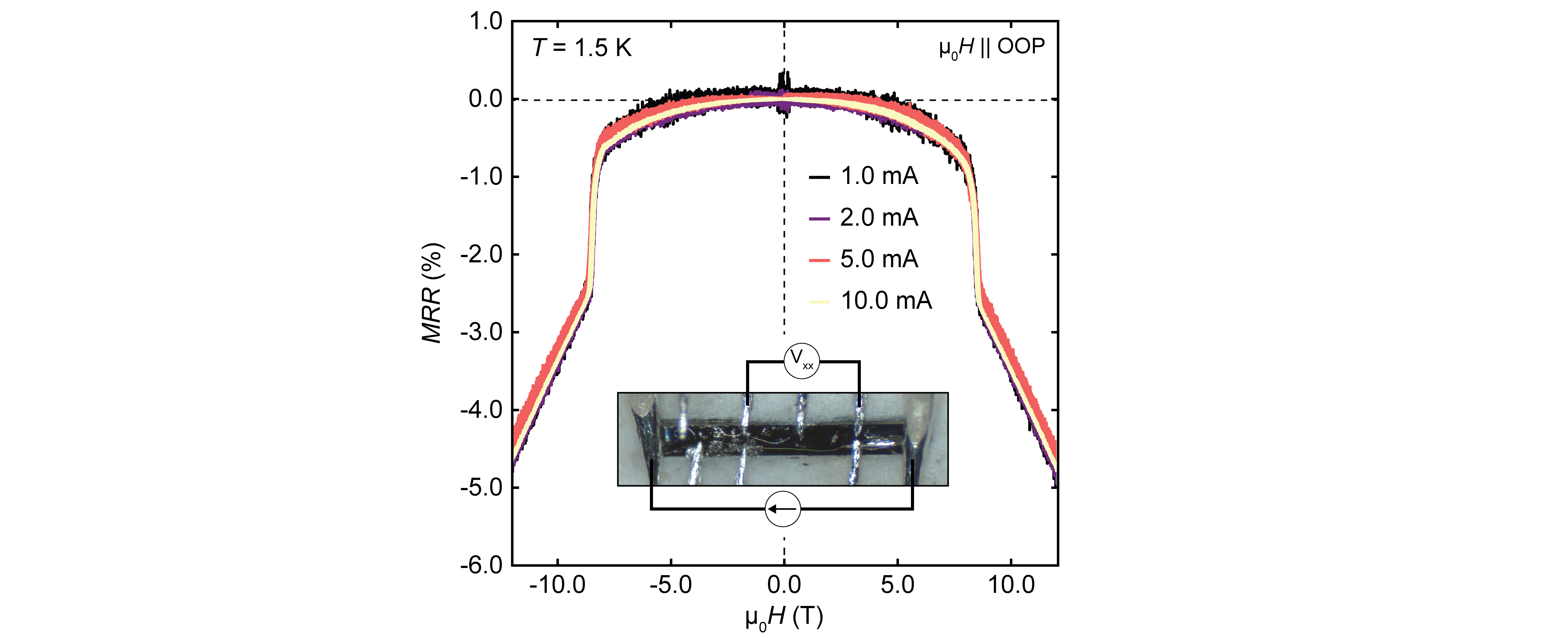}% Here is how to import EPS art
\centering
\caption[Current dependence of magnetotransport in TaFe\textsubscript{1.14}Te\textsubscript{3}]{\label{fig:figures19} Current dependence of magnetotransport in TaFe\textsubscript{1.14}Te\textsubscript{3}. Magnetoresistance ratio, \textit{MRR} = [\textit{R}(\textit{H})-\textit{R}(\textit{H}=0)]/\textit{R}(\textit{H}=0)$\times$100, versus magnetic field at various source currents for fields parallel to the out-of-plane direction. Both forward and backward traces are shown. The traces were all taken at 1.5 K. The dashed black lines denote the zero crossings. The inset is an optical image of the device along with the measurement configuration.}
\end{figure*}
%%%%%%%%%%%%%%%%%%%%%%%%%%%%%%%%%%%%%%%%%%%%%%%%%%%%%%%%%%%%%%%%%%%%%%%%%%%%%%%%%%%%%%%%%%%%%%%%%%%%%%%%%%%%
%%%%%%%%%%%%%%%%%%%%%%%%%%%%%%%%%%%%%%%%%%%%%%%%%%%%%%%%%%%%%%%%%%%%%%%%%%%%%%%%%%%%%%%%%%%%%%%%%%%%%%%%%%%%

%%%%%%%%%%%%%%%%%%%%%%%%%%%%%%%%%%%%%%%%%%%%%%%%%%%%%%%%%%%%%%%%%%%%%%%%%%%%%%%%%%%%%%%%%%%%%%%%%%%%%%%%%%%%
%Figure s20
%%%%%%%%%%%%%%%%%%%%%%%%%%%%%%%%%%%%%%%%%%%%%%%%%%%%%%%%%%%%%%%%%%%%%%%%%%%%%%%%%%%%%%%%%%%%%%%%%%%%%%%%%%%%
\begin{figure*}[t]
\includegraphics[width=6.9in]{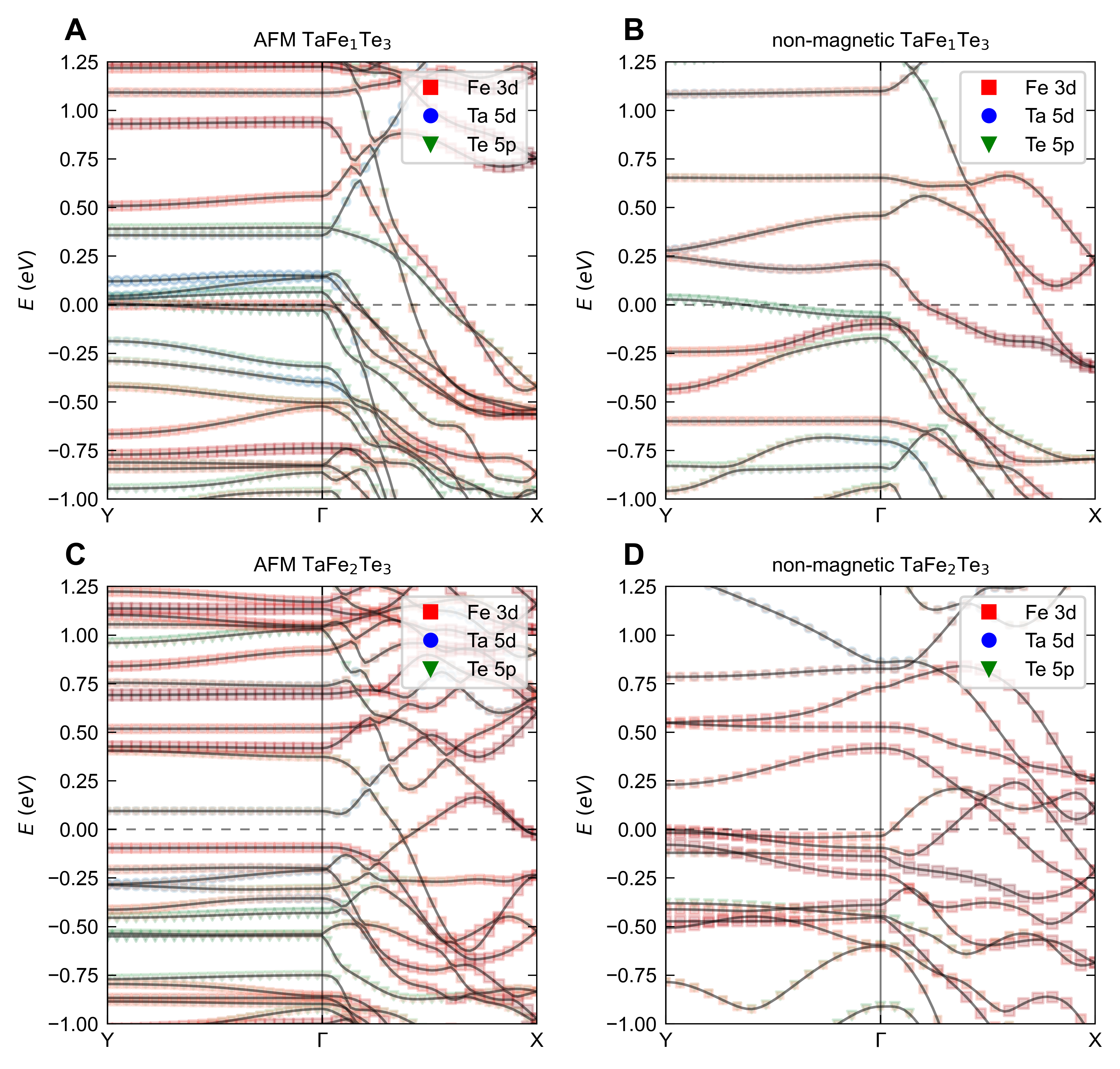}% Here is how to import EPS art
\centering
\caption[ Calculated band structures for TaFeTe\textsubscript{3} and TaFe\textsubscript{2}Te\textsubscript{3}]{\label{fig:figures20} Calculated band structures for TaFeTe\textsubscript{3} and TaFe\textsubscript{2}Te\textsubscript{3}. A, B) Element-resolved band structure calculations for TaFeTe\textsubscript{3} in the antiferromagnetic (A) and non-magnetic (B) configurations. Red squares, blue circles, and green triangles correspond to Fe 3d orbitals, Ta 5d orbitals, and Te 5p orbitals. C, D). Element-resolved band structure calculations for TaFe\textsubscript{2}Te\textsubscript{3} in the antiferromagnetic (C) and non-magnetic (D) state. Red squares, blue circles, and green triangles correspond to Fe 3d orbitals, Ta 5d orbitals, and Te 5p orbitals.}
\end{figure*}
%%%%%%%%%%%%%%%%%%%%%%%%%%%%%%%%%%%%%%%%%%%%%%%%%%%%%%%%%%%%%%%%%%%%%%%%%%%%%%%%%%%%%%%%%%%%%%%%%%%%%%%%%%%%
%%%%%%%%%%%%%%%%%%%%%%%%%%%%%%%%%%%%%%%%%%%%%%%%%%%%%%%%%%%%%%%%%%%%%%%%%%%%%%%%%%%%%%%%%%%%%%%%%%%%%%%%%%%%

%%%%%%%%%%%%%%%%%%%%%%%%%%%%%%%%%%%%%%%%%%%%%%%%%%%%%%%%%%%%%%%%%%%%%%%%%%%%%%%%%%%%%%%%%%%%%%%%%%%%%%%%%%%%
%Figure s21
%%%%%%%%%%%%%%%%%%%%%%%%%%%%%%%%%%%%%%%%%%%%%%%%%%%%%%%%%%%%%%%%%%%%%%%%%%%%%%%%%%%%%%%%%%%%%%%%%%%%%%%%%%%%
\begin{figure*}[t]
\includegraphics[width=6.9in]{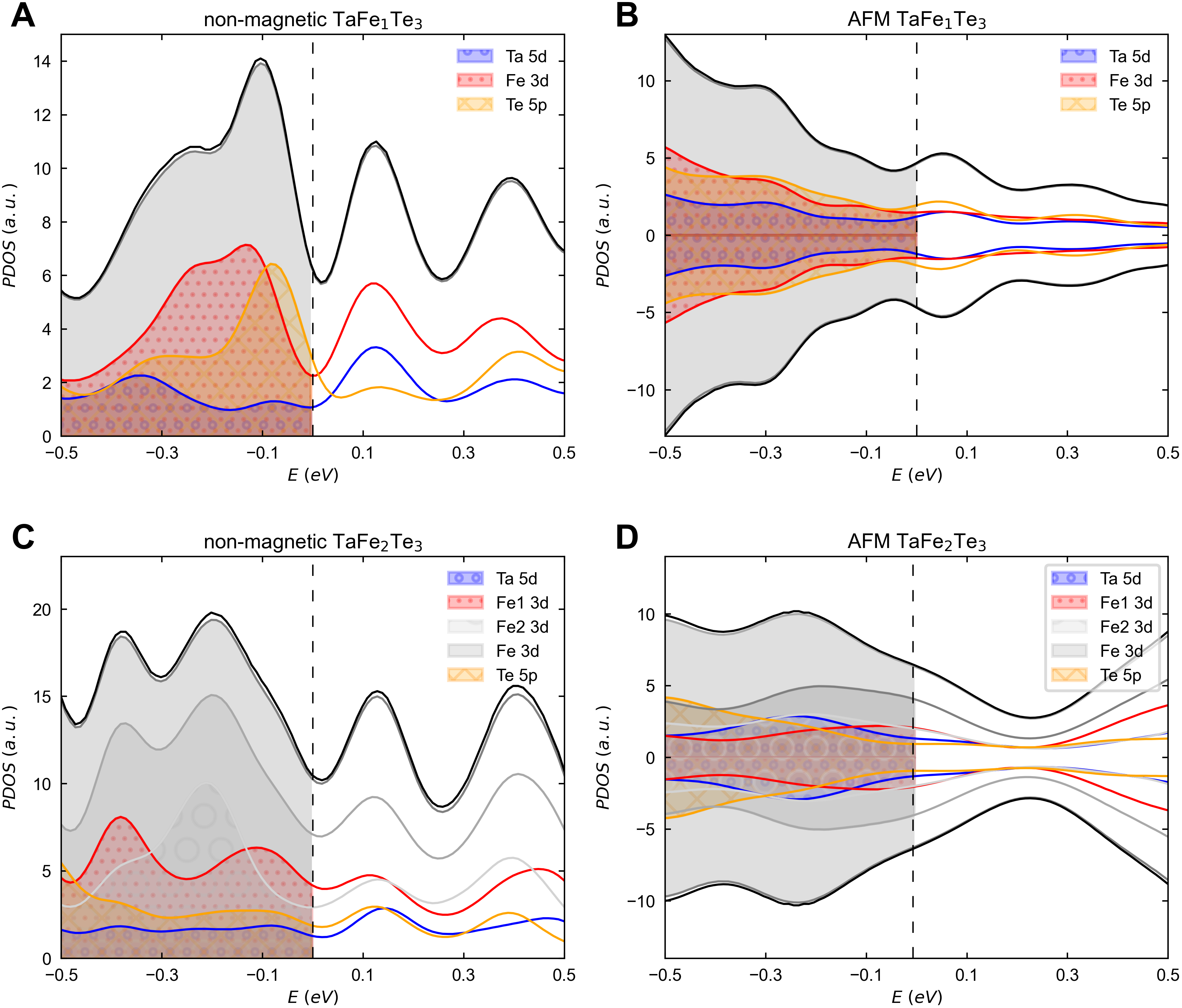}% Here is how to import EPS art
\centering
\caption[Element-resolved calculated partial density of states (PDOS) for TaFe\textsubscript{1+\textit{y}}Te\textsubscript{3}]{\label{fig:figures21} Element-resolved calculated partial density of states (PDOS) for TaFe\textsubscript{1+\textit{y}}Te\textsubscript{3}. A, B) Element-resolved partial density of states (PDOS) calculated for TaFeTe\textsubscript{3} in the non-magnetic (A) and antiferromagnetic (B) configuration. C, D) Element-resolved partial density of states (PDOS) calculated for TaFe\textsubscript{2}Te\textsubscript{3} in the non-magnetic (C) and antiferromagnetic (D) configuration. }
\end{figure*}
%%%%%%%%%%%%%%%%%%%%%%%%%%%%%%%%%%%%%%%%%%%%%%%%%%%%%%%%%%%%%%%%%%%%%%%%%%%%%%%%%%%%%%%%%%%%%%%%%%%%%%%%%%%%
%%%%%%%%%%%%%%%%%%%%%%%%%%%%%%%%%%%%%%%%%%%%%%%%%%%%%%%%%%%%%%%%%%%%%%%%%%%%%%%%%%%%%%%%%%%%%%%%%%%%%%%%%%%%

%%%%%%%%%%%%%%%%%%%%%%%%%%%%%%%%%%%%%%%%%%%%%%%%%%%%%%%%%%%%%%%%%%%%%%%%%%%%%%%%%%%%%%%%%%%%%%%%%%%%%%%%%%%%
%Figure s22
%%%%%%%%%%%%%%%%%%%%%%%%%%%%%%%%%%%%%%%%%%%%%%%%%%%%%%%%%%%%%%%%%%%%%%%%%%%%%%%%%%%%%%%%%%%%%%%%%%%%%%%%%%%%
\begin{figure*}[t]
\includegraphics[width=6.9in]{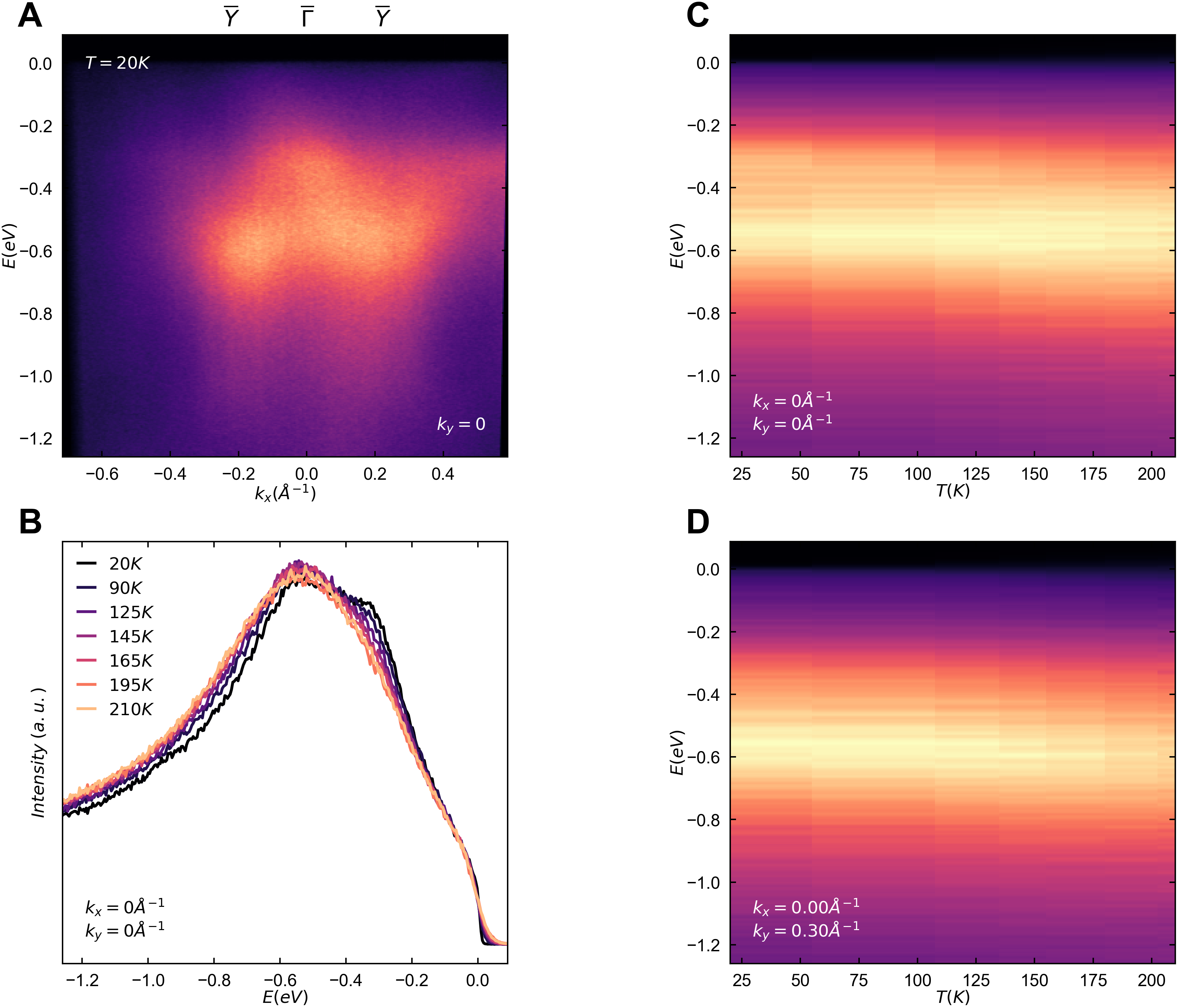}% Here is how to import EPS art
\centering
\caption[Temperature dependence of ARPES spectra]{\label{fig:figures22} Temperature dependence of ARPES spectra. A) ARPES spectrum along $\bar{Y}-\bar{\Gamma}-\bar{Y}$ path at 20 K. The purple and yellow colors represent the lowest and highest photoelectron intensities, respectively. B) EDCs at various temperatures at k\textsubscript{x} = 0 \r{A}\textsuperscript{-1}, showing a two-peak structure corresponding to two bands. The peak near the Fermi level shows enhancement of spectral intensity at low temperatures, but negligible above $\approx$125 K.  C) Change of the energy position at k\textsubscript{x} = 0 \r{A}\textsuperscript{-1} ($\bar{\Gamma}$) with temperature. The top band slightly shifted downward with increasing temperature and no measurable band shift is observed for the bottom band, except for the increase of broadening. D) Change of the energy position at k\textsubscript{x} = 0.31 \r{A}\textsuperscript{-1} ($\bar{Y}$) with temperature. The band appears to shift slightly downward with increasing temperature across the magnetic transition. For (C) and (D), the purple and yellow colors represent the lowest and highest photoelectron intensities, respectively.}
\end{figure*}
%%%%%%%%%%%%%%%%%%%%%%%%%%%%%%%%%%%%%%%%%%%%%%%%%%%%%%%%%%%%%%%%%%%%%%%%%%%%%%%%%%%%%%%%%%%%%%%%%%%%%%%%%%%%
%%%%%%%%%%%%%%%%%%%%%%%%%%%%%%%%%%%%%%%%%%%%%%%%%%%%%%%%%%%%%%%%%%%%%%%%%%%%%%%%%%%%%%%%%%%%%%%%%%%%%%%%%%%%

%%%%%%%%%%%%%%%%%%%%%%%%%%%%%%%%%%%%%%%%%%%%%%%%%%%%%%%%%%%%%%%%%%%%%%%%%%%%%%%%%%%%%%%%%%%%%%%%%%%%%%%%%%%%
%Figure s23
%%%%%%%%%%%%%%%%%%%%%%%%%%%%%%%%%%%%%%%%%%%%%%%%%%%%%%%%%%%%%%%%%%%%%%%%%%%%%%%%%%%%%%%%%%%%%%%%%%%%%%%%%%%%
\begin{figure*}[t]
\includegraphics[width=6.9in]{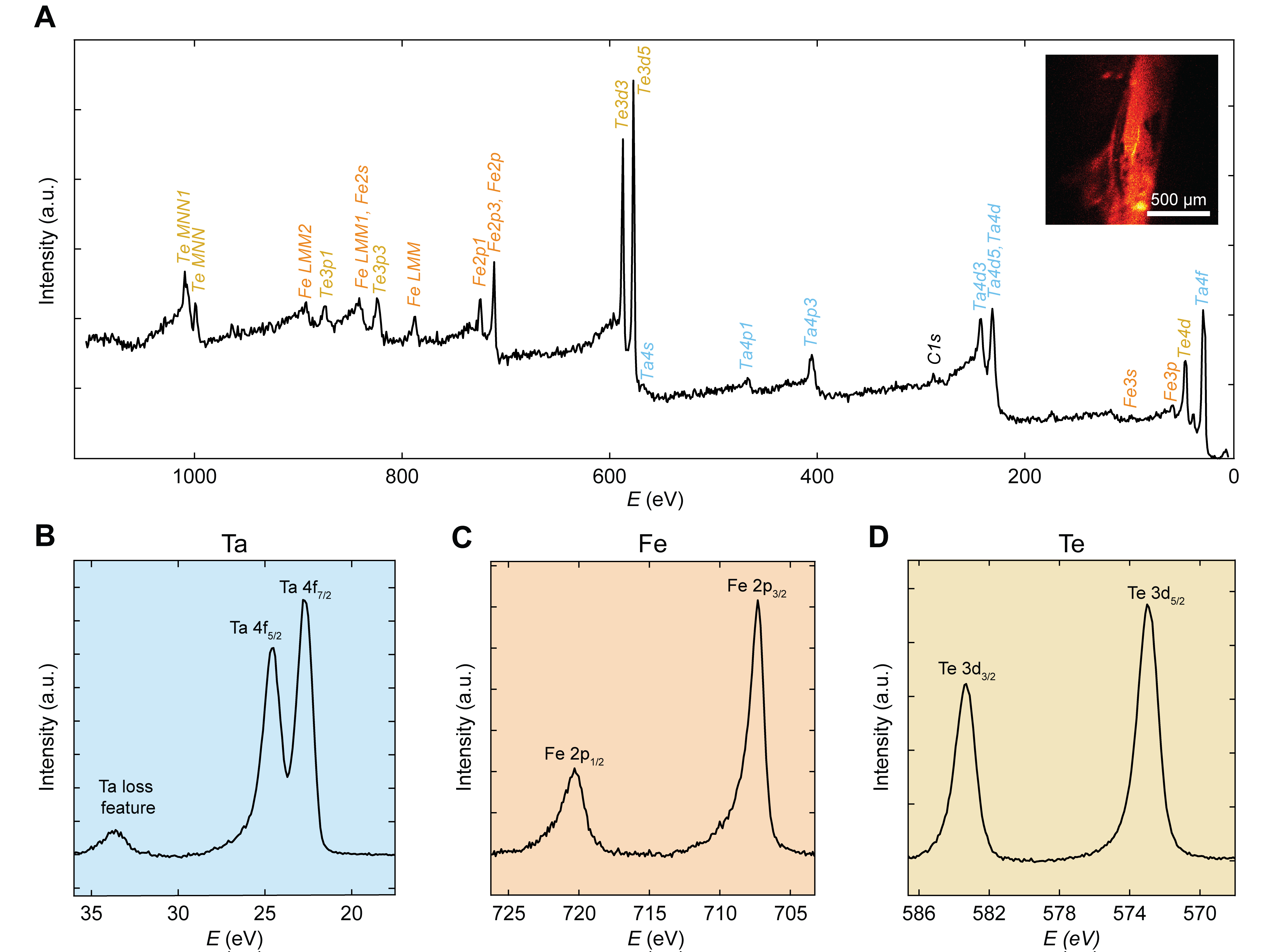}% Here is how to import EPS art
\centering
\caption[X-ray photoelectron spectroscopy on bulk TaFe\textsubscript{1.14}Te\textsubscript{3}]{\label{fig:figures23} X-ray photoelectron spectroscopy on bulk TaFe\textsubscript{1.14}Te\textsubscript{3}. A) Intensity versus binding energy survey of a bulk sample taken at 298 K with a beam diameter of 100 {\textmu}m and a source power of 25 W. Ta, Fe, and Te spectral regions are noted with blue, orange, and yellow, respectively. B-D) Intensity versus binding energy focused on expected Ta (B), Fe (C), and Te (D) peak regions.}
\end{figure*}
%%%%%%%%%%%%%%%%%%%%%%%%%%%%%%%%%%%%%%%%%%%%%%%%%%%%%%%%%%%%%%%%%%%%%%%%%%%%%%%%%%%%%%%%%%%%%%%%%%%%%%%%%%%%
%%%%%%%%%%%%%%%%%%%%%%%%%%%%%%%%%%%%%%%%%%%%%%%%%%%%%%%%%%%%%%%%%%%%%%%%%%%%%%%%%%%%%%%%%%%%%%%%%%%%%%%%%%%%
\setlength{\extrarowheight}{2mm}
\renewcommand\thetable{S\arabic{table}}    
\setcounter{figure}{0} 
%%%%%%%%%%%%%%%%%%%%%%%%%%%%%%%%%%%%%%%%%%%%%%%%%%%%%%%%%%%%%%%%%%%%%%%%%%%%%%%%%%%%%%%%%%%%%%%%%%%%%%%%%%%%
%Table s1
%%%%%%%%%%%%%%%%%%%%%%%%%%%%%%%%%%%%%%%%%%%%%%%%%%%%%%%%%%%%%%%%%%%%%%%%%%%%%%%%%%%%%%%%%%%%%%%%%%%%%%%%%%%%
\begin{table*}[t]
\begin{ruledtabular}
\begin{tabular}{cccc}
Temperature&300 K&200 K&100 K\\[2mm] \hline
Empirical Formula&TaFe\textsubscript{1.14}Te\textsubscript{3}&TaFe\textsubscript{1.12}Te\textsubscript{3}&TaFe\textsubscript{1.12}Te\textsubscript{3}\\
Formula weight&627.42&626.02&626.58\\
Temperature&300.01(10) K&200.00(10) K&100.00(10) K\\
Wavelength&0.71073 \r{A}&0.71073 \r{A}&0.71073 \r{A}\\
Crystal system&Monoclinic&Monoclinic&Monoclinic\\
Space group&\textit{P}2\textsubscript{1}/\textit{m}&\textit{P}2\textsubscript{1}/\textit{m}&\textit{P}2\textsubscript{1}/\textit{m}\\
\textit{a}&7.4171(3)\r{A}&7.4076(3)\r{A}&7.4105(4)\r{A}\\
\textit{\textalpha}&90\textsuperscript{$\circ$}&90\textsuperscript{$\circ$}&90\textsuperscript{$\circ$}\\
\textit{b}&3.63280(10)\r{A}&3.62530(10)\r{A}&3.61940(10)\r{A}\\
\textit{\textbeta}&109.129(5)\textsuperscript{$\circ$}&109.175(4)\textsuperscript{$\circ$}&109.189(6)\textsuperscript{$\circ$}\\
\textit{c}&9.9740(4)\r{A}&9.9661(4)\r{A}&9.9626(5)\r{A}\\
\textit{\textgamma}&90\textsuperscript{$\circ$}&90\textsuperscript{$\circ$}&90\textsuperscript{$\circ$}\\
Volume&253.908(18)\r{A}\textsuperscript{3}&252.789(17)\r{A}\textsuperscript{3}&252.37(2)\r{A}\textsuperscript{3}\\
Z&2&2&2\\
Density (calculated)&8.207 g/cm\textsuperscript{3}&8.225 g/cm\textsuperscript{3}&8.246 g/cm\textsuperscript{3}\\
Absorption coefficient&41.478 mm\textsuperscript{-1}&41.593 mm\textsuperscript{-1}&41.690 mm\textsuperscript{-1}\\
F(000)&517&516&516\\
Crystal size&0.239$\times$0.049$\times$0.019 mm\textsuperscript{3}&0.239$\times$0.049$\times$0.019 mm\textsuperscript{3}&0.239$\times$0.049$\times$0.019 mm\textsuperscript{3}\\
\texttheta collection range&2.161\textsuperscript{$\circ$} to 28.400\textsuperscript{$\circ$}&2.164\textsuperscript{$\circ$} to 28.402\textsuperscript{$\circ$}&2.164\textsuperscript{$\circ$} to 28.539\textsuperscript{$\circ$}\\
Index ranges&-9$\leq$\textit{h}$\leq$9,-4$\leq$\textit{k}$\leq$4,-13$\leq$\textit{l}$\leq$12&-9$\leq$\textit{h}$\leq$9,-4$\leq$\textit{k}$\leq$4,-13$\leq$\textit{l}$\leq$12&-9$\leq$\textit{h}$\leq$9,-4$\leq$\textit{k}$\leq$4,-13$\leq$\textit{l}$\leq$12\\
Reflections collected&2966&2950&2962\\
Independent reflection&681 [\textit{R}\textsubscript{int}=0.0410]&674 [\textit{R}\textsubscript{int}=0.0428]&682 [\textit{R}\textsubscript{int}=0.0498]\\
Completeness to \texttheta=25.242\textsuperscript{$\circ$}&100\%&100\%&100\%\\
Refinement method&Full-matrix least-squares on F\textsuperscript{2}&Full-matrix least-squares on F\textsuperscript{2}&Full-matrix least-squares on F\textsuperscript{2}\\
Data/restraints/parameters&681/0/35&674/0/35&682/0/35\\
Goodness-of-fit&1.082&1.121&1.047\\
Final \textit{R} indices[\textit{I}$>$2\textsigma(\textit{I})]&\textit{R}\textsubscript{obs}=0.0261, \textit{wR}\textsubscript{obs}=0.0601&\textit{R}\textsubscript{obs}=0.0267, \textit{wR}\textsubscript{obs}=0.0578&\textit{R}\textsubscript{obs}=0.0284, \textit{wR}\textsubscript{obs}=0.0627\\
\textit{R} indices [all data]&\textit{R}\textsubscript{all}=0.0291, \textit{wR}\textsubscript{all}=0.0624&\textit{R}\textsubscript{all}=0.0283, \textit{wR}\textsubscript{all}=0.0587&\textit{R}\textsubscript{all}=0.0299, \textit{wR}\textsubscript{all}=0.0638\\
Extinction coefficient&0.0015(3)&0.0024(3)&0.0019(3)\\
Largest diff. peak and hole&1.725 \textit{e}$\cdot$\r{A}\textsuperscript{-3} and -2.003 \textit{e}$\cdot$\r{A}\textsuperscript{-3}&2.093 \textit{e}$\cdot$\r{A}\textsuperscript{-3} and -2.263 \textit{e}$\cdot$\r{A}\textsuperscript{-3}&2.136 \textit{e}$\cdot$\r{A}\textsuperscript{-3} and -2.824 \textit{e}$\cdot$\r{A}\textsuperscript{-3}\\
\end{tabular}
\end{ruledtabular}
\caption[Crystallographic parameters determined through SCXRD for TaFe\textsubscript{1.14}Te\textsubscript{3} at various temperatures]{\label{tab:tables1} Crystallographic parameters determined through SCXRD for TaFe\textsubscript{1.14}Te\textsubscript{3} at various temperatures. \textit{R}=\textSigma$|$$|$\textit{F}\textsubscript{0}$|$-$|$\textit{F}\textsubscript{C}$|$ $|$/\textSigma$|$\textit{F}\textsubscript{0}$|$, \textit{wR} = 
 \{{\textSigma[\textit{w}($|$\textit{F}\textsubscript{0}$|$\textsuperscript{2} - $|$\textit{F}\textsubscript{C}$|$\textsuperscript{2})\textsuperscript{2}]/\textSigma[\textit{w}($|$\textit{F}\textsubscript{0}$|$\textsuperscript{4})}]\}\textsuperscript{1/2} and \textit{w}=1/[\textsigma\textsuperscript{2}(\textit{F}\textsubscript{0}\textsuperscript{2})+(0.021\textit{P})\textsuperscript{2}+1.7407\textit{P}], where \textit{P}=(\textit{F}\textsubscript{0}\textsuperscript{2}+2\textit{F}\textsubscript{C}\textsuperscript{2})/3.}
\end{table*}
%%%%%%%%%%%%%%%%%%%%%%%%%%%%%%%%%%%%%%%%%%%%%%%%%%%%%%%%%%%%%%%%%%%%%%%%%%%%%%%%%%%%%%%%%%%%%%%%%%%%%%%%%%%%
%%%%%%%%%%%%%%%%%%%%%%%%%%%%%%%%%%%%%%%%%%%%%%%%%%%%%%%%%%%%%%%%%%%%%%%%%%%%%%%%%%%%%%%%%%%%%%%%%%%%%%%%%%%%
%%%%%%%%%%%%%%%%%%%%%%%%%%%%%%%%%%%%%%%%%%%%%%%%%%%%%%%%%%%%%%%%%%%%%%%%%%%%%%%%%%%%%%%%%%%%%%%%%%%%%%%%%%%%
%Table s2
%%%%%%%%%%%%%%%%%%%%%%%%%%%%%%%%%%%%%%%%%%%%%%%%%%%%%%%%%%%%%%%%%%%%%%%%%%%%%%%%%%%%%%%%%%%%%%%%%%%%%%%%%%%%
\begin{table*}[t]
\begin{ruledtabular}
\begin{tabular}{ccccccc}
 \multicolumn{7}{c}{Growth Batches}\\[2mm]
 &EJT-001&RJ-1-30&SJH-164&SJH-171&SJH-208&Combined\\[2mm] \hline
 Ta&1.00$\pm$0.08&0.88$\pm$0.04&0.89$\pm$0.06&0.85$\pm$0.03&0.87$\pm$0.02&0.90$\pm$0.06\\[2mm] \hline
 Fe&1.21$\pm$0.05&1.19$\pm$0.05&1.16$\pm$0.06&1.26$\pm$0.25&1.22$\pm$0.03&1.21$\pm$0.04\\[2mm] \hline
 Te&3.00&3.00&3.00&3.00&3.00&3.00\\
\end{tabular}
\end{ruledtabular}
\caption[Extracted chemical compositions for multiple TaFe\textsubscript{1.14}Te\textsubscript{3} growth batches]{\label{tab:tables2} Extracted chemical compositions for multiple TaFe\textsubscript{1.14}Te\textsubscript{3} growth batches. Relative concentration of Ta, Fe, and Te normalized to the measured Te concentration for each growth batch. The last column is the average concentration of each element between all growth batches. Error bars represent the standard deviation. All chemical concentrations were extracted from SEM/EDX data.}
\end{table*}
%%%%%%%%%%%%%%%%%%%%%%%%%%%%%%%%%%%%%%%%%%%%%%%%%%%%%%%%%%%%%%%%%%%%%%%%%%%%%%%%%%%%%%%%%%%%%%%%%%%%%%%%%%%%
%%%%%%%%%%%%%%%%%%%%%%%%%%%%%%%%%%%%%%%%%%%%%%%%%%%%%%%%%%%%%%%%%%%%%%%%%%%%%%%%%%%%%%%%%%%%%%%%%%%%%%%%%%%%
%%%%%%%%%%%%%%%%%%%%%%%%%%%%%%%%%%%%%%%%%%%%%%%%%%%%%%%%%%%%%%%%%%%%%%%%%%%%%%%%%%%%%%%%%%%%%%%%%%%%%%%%%%%%
%Table s3
%%%%%%%%%%%%%%%%%%%%%%%%%%%%%%%%%%%%%%%%%%%%%%%%%%%%%%%%%%%%%%%%%%%%%%%%%%%%%%%%%%%%%%%%%%%%%%%%%%%%%%%%%%%%
\begin{table*}[t]
\begin{ruledtabular}
\begin{tabular}{ccccc}
 \multicolumn{5}{c}{TaFeTe\textsubscript{3}}\\
 \multicolumn{2}{c}{Non-magnetic}&\multicolumn{3}{c}{Antiferromagnetic}\\ 
 Element&Charge&Element&Charge&Mag. Moment\\
 Ta&0.3&Ta&0.3&$\pm$0.2\\
 Fe1&0.4&Fe1&0.5&2.2\\
 Fe2&\textemdash&Fe2&\textemdash&\textemdash\\
 Te&-0.1&Te&0.0&0.0\\[2mm] \hline

 \multicolumn{5}{c}{TaFe\textsubscript{1.25}Te\textsubscript{3}}\\
 \multicolumn{2}{c}{Non-magnetic}&\multicolumn{3}{c}{Antiferromagnetic}\\
 Element&Charge&Element&Charge&Mag. Moment\\
 Ta&0.2&Ta&0.3&$\pm$0.2\\
 Fe1&0.4&Fe1&0.5&2.2\\
 Fe2&0.4&Fe2&0.5&2.3\\
 Te&0.0&Te&0.0&0.0\\[2mm] \hline
 
 \multicolumn{5}{c}{TaFe\textsubscript{2}Te\textsubscript{3}}\\
 \multicolumn{2}{c}{Non-magnetic}&\multicolumn{3}{c}{Antiferromagnetic}\\
 Element&Charge&Element&Charge&Mag. Moment\\
 Ta&0.2&Ta&0.1&$\pm$0.1\\
 Fe1&0.1&Fe1&0.4&2.3\\
 Fe2&0.1&Fe2&0.3&2.4\\
 Te&0.0&Te&-0.1&0.0\\
 
\end{tabular}
\end{ruledtabular}
\caption[Calculated Löwdin charge and magnetic moments for TaFe\textsubscript{1+\textit{y}}Te\textsubscript{3}]{\label{tab:tables3} Calculated Löwdin charge and magnetic moments for TaFe\textsubscript{1+\textit{y}}Te\textsubscript{3}. Calculated charge and magnetic moment per atom (Ta, Fe1, Fe2, and Te) for TaFeTe\textsubscript{3}, TaFe\textsubscript{1.25}Te\textsubscript{3}, and TaFe\textsubscript{2}Te\textsubscript{3}. Note that due to the arbitrary nature of atomic orbitals used for projection, the Löwdin charges do not necessarily sum to zero.}
\end{table*}
%%%%%%%%%%%%%%%%%%%%%%%%%%%%%%%%%%%%%%%%%%%%%%%%%%%%%%%%%%%%%%%%%%%%%%%%%%%%%%%%%%%%%%%%%%%%%%%%%%%%%%%%%%%%
%%%%%%%%%%%%%%%%%%%%%%%%%%%%%%%%%%%%%%%%%%%%%%%%%%%%%%%%%%%%%%%%%%%%%%%%%%%%%%%%%%%%%%%%%%%%%%%%%%%%%%%%%%%%
\end{document}